\documentclass{emulateapj}
\newcommand{\chn}{{\it Chandra}}
\newcommand{\swf}{{\it Swift}}

\usepackage{graphicx}
\usepackage{longtable}
\usepackage{mdwlist}
\usepackage{natbib}
\usepackage{multirow}
\usepackage{upgreek}
\usepackage[colorlinks=true, pdfstartview=FitV, linkcolor=blue,citecolor=blue, urlcolor=blue]{hyperref}
\usepackage{wasysym}
\usepackage{txfonts}
\usepackage{gensymb}
\usepackage{morefloats}
\usepackage{threeparttable}
\usepackage{rotating}
\usepackage{color}

\shorttitle{3CR unidentified sources}
\shortauthors{V. Missaglia et al. 2021}

\begin{document}
\title{Hidden treasures in the unknown 3CR extragalactic radio sky:\\ a multi-wavelength approach}

\author{
V. Missaglia\altaffilmark{1,2,3},
F. Massaro\altaffilmark{1,2,3}, 
E. Liuzzo\altaffilmark{4},
A. Paggi\altaffilmark{1,3},
R. P. Kraft\altaffilmark{5},
W. R. Forman\altaffilmark{5},
A. Jimenez-Gallardo\altaffilmark{1,2,3,6},
J. P. Madrid\altaffilmark{7},
F. Ricci\altaffilmark{8,9},
C. Stuardi\altaffilmark{8,9},
B. J. Wilkes\altaffilmark{5},
S. A. Baum\altaffilmark{10},
C. P. O'Dea\altaffilmark{10},
J. Kuraszkiewicz\altaffilmark{5},
G. R. Tremblay\altaffilmark{5},
A. Maselli\altaffilmark{11},
A. Capetti\altaffilmark{2},
E. Sani\altaffilmark{6},
B. Balmaverde\altaffilmark{2}
\&
D. E. Harris\altaffilmark{5,+}
}

\altaffiltext{1}{Dipartimento di Fisica, Universit\`a degli Studi di Torino, via Pietro Giuria 1, I-10125 Torino, Italy.}
\altaffiltext{2}{INAF-Osservatorio Astrofisico di Torino, via Osservatorio 20, I-10025 Pino Torinese, Italy.}
\altaffiltext{3}{INFN-Istituto Nazionale di Fisica Nucleare, Sezione di Torino, I-10125 Torino, Italy.}
\altaffiltext{4}{INAF-IRA, via Piero Gobetti 101, I-40129, Bologna, Italy.}
\altaffiltext{5}{Center for Astrophysics $\mid$ Harvard \& Smithsonian, 60 Garden Street, Cambridge, MA 02138, USA.}
\altaffiltext{6}{European Southern Observatory, Alonso de C\'ordova 3107, Vitacura, Casilla 19001, Santiago de Chile, Chile.}
\altaffiltext{7}{University of Texas - Rio Grande Valley, One West University Blvd, Brownsville, TX 78520, USA.}
\altaffiltext{8}{Dipartimento di Fisica e Astronomia, Universit\`a di Bologna, via Piero Gobetti 93/2, I-40129 Bologna, Italy.}
\altaffiltext{9}{INAF - Osservatorio di Astrofisica e Scienza dello Spazio di Bologna, Via Piero Gobetti, 93/3, I–40129 Bologna, Italy.}
\altaffiltext{10}{University of Manitoba,  Dept. of Physics and Astronomy, Winnipeg, MB R3T 2N2, Canada.}
\altaffiltext{11}{Space Science Data Center - Agenzia Spaziale Italiana, Via del Politecnico, snc, I-00133, Roma, Italy.}
\altaffiltext{+}{Dan Harris passed away on December 6th, 2015. His career spanned much of the history of radio and X-ray astronomy. His passion, insight, and contributions will always be remembered. A significant fraction of this work is one of his last efforts.}

\begin{abstract} 
We present the analysis of multi-wavelength observations of seven extragalactic radio sources, listed as unidentified in the Third Cambridge Revised Catalog (3CR). X-ray observations, performed during \chn\ Cycle 21, were compared to VLA, WISE and Pan-STARRS observations in the radio, infrared and optical bands, respectively. All sources in this sample lack a clear optical counterpart, and are thus missing their redshift and optical classification. In order to confirm the X-ray and infrared radio counterparts of core and extended components, here we present for the first time radio maps obtained manually reducing VLA archival data. As in previous papers on the \chn\ X-ray snapshot campaign, we report X-ray detections of radio cores and two sources, out of the seven presented here, are found to be members of galaxy clusters. For these two cluster sources (namely, 3CR\,409 and 3CR\,454.2), we derived surface brightness profiles in four directions. For all seven sources, we measured X-ray intensities of the radio sources and we also performed standard X-ray spectral analysis for the four sources (namely, 3CR\,91, 3CR\,390, 3CR\,409 and 3CR\,428) with the brightest nuclei (more than 400 photons in the 2\arcsec\ nuclear region). We also detected extended X-ray emission around 3CR\,390 and extended X-ray emission associated with the northern jet of 3CR\,158. This paper represents the first attempt to give a multi-wavelength view of the unidentified radio sources listed in the 3CR catalog.
\end{abstract}

\section{Introduction}\label{sec:intro}
The Third Cambridge Catalog of radio sources \citep[3C,][]{1959MmRAS..68...37E} performed at 159 MHz, and its revised releases at 178 MHz (3CR: \citealt{1962MmRAS..68..163B}; 3CRR: \citealt{1983MNRAS.204..151L} and the \citealt{1985PASP...97..932S} update) are paramount low-frequency radio catalogs for studying radio loud, active galactic nuclei (AGNs) and their environments at all scales \citep[see e.g.][]{2012ARA&A..50..455F,2012ApJ...749...19K,2020MNRAS.492.3156L}. To a great degree, the success of this catalog is due to the fact that its latest revision (3CRR) has a flux limit of 9 Jy at 178 MHz and { it represents} a statistically complete sample of the most powerful radio galaxies, including a variety of extended radio morphologies, optical classes, and environmental properties.

On the basis of the 3CR radio observations at 178 MHz, \citet{1974MNRAS.167P..31F} proposed a classification for radio sources based on the relative position of regions of high and low surface brightness in their extended structures, distinguishing between FRI, i.e. edge-darkened, and FRII, i.e. edge-brightened, types. Since 1974, a lot more has been learned on the FRI/FRII dichotomy, as reported in \citet{1984AJ.....89..979B,1995ApJ...451...88B,2000A&A...355..873C,2019MNRAS.488.2701M}. Between '80s and '90s, an additional classification was proposed, on the basis of the relative intensity of high and low excitation lines in the optical spectra \citep{1979MNRAS.188..111H,1994ASPC...54..201L}. Two populations of radio galaxies were then defined: high-excitation radio galaxies, or HERGs, and low-excitation radio galaxies, or LERGs. These two classes are believed to represent intrinsically different types of objects, since they show different accretion rates \citep{2002A&A...394..791C,2009MNRAS.396.1929H,2012MNRAS.421.1569B}, host galaxies and redshift evolution \citep{2016MNRAS.460....2P}.


In the last three decades, several photometric and spectroscopic surveys of 3CR radio sources have been carried out. For example, using the Hubble Space Telescope (HST), the 3CR catalog has been observed in the near-ultraviolet \citep{2002ApJS..139..411A}, optical \citep{1996ApJS..107..621D,1997ApJS..112..415M,1999ApJS..122...81M,2008ApJS..175..423P,2009ApJS..183..278T,2017FrASS...4...52R} and near-infrared \citep{2006ApJS..164..307M,2010ApJ...725.2426B}. \citet{2009A&A...495.1033B} carried out an optical spectroscopic survey of 3CR  radio galaxies with the Telescopio Nazionale Galileo. More recently, \citet{2019A&A...632A.124B} presented Very Large Telescope/Multi Unit Spectroscopic Explorer (VLT/MUSE) observations of 20 low-z 3CR radio galaxies. Finally, Very Long Baseline Array (VLBA) data for several 3CR objects at redshifts $z<0.2$ were also obtained \citep[see e.g.,][and references therein]{2001ApJ...552..508G,2009A&A...505..509L}.

In X-rays, most of the 3CR extragalactic radio sources were observed with \chn, XMM-Newton and \swf\ \citep[see e.g.][]{2001MNRAS.326.1127W,2005MNRAS.358..843H,2007ApJ...659.1008K}. Until  Cycle  9  the \chn\ archive  covered  only $\sim 60\%$ of  the  3CR  extragalactic  sample,  while the other X-ray telescopes, such as XMM-Newton, had obtained data for less than 1/3 of the entire catalog.  

In 2008, the 3C \chn\ snapshot campaign began, aiming to detect X-ray emission in extragalactic radio sources arising from jets and hotspots, to determine their X-ray emission processes on a firm statistical basis, and to study the nuclear emission of the host galaxies \citep{2010ApJ...714..589M}. In the \chn\ archive, 150 out of 298 3CR extragalactic sources were already present before the beginning of the survey. We observed 123 3CR sources, and X-ray emission has been detected in 119 out of 122 radio cores, in addition to the discovery of the X-ray counterpart for eight jet knots, 23 hotspots, with marginal detection for another nine, and 17 radio lobes \citep[see][for the latest run of Chandra observations]{2020ApJS..250....7J} . Diffuse X-ray emission was also detected around several 3C radio sources, potentially associated with either their radio lobes or radiation arising from the intergalactic medium (IGM) when harbored in galaxy clusters/groups \citep[see e.g. 3C\,17, 3C\,196.1, 3C\,89, and 3C\,187][respectively]{2018ApJS..238...31M,2018ApJ...867...35R,2016MNRAS.458..681D,2020arXiv201211610P}. Some of the most interesting 3CR sources have been investigated in more detail with \chn\ follow up observations, such as 3C\,171, 3C\,105 and 3C\,305 to name a few \citep[see e.g.][respectively]{2010MNRAS.401.2697H,2012MNRAS.419.2338O,2012MNRAS.424.1774H}. Additional \chn\ X-ray observations, restricted to the 3CRR catalog, have been also carried out in parallel during the last decade \citep[see e.g.][]{2013ApJ...773...15W}.

During our campaign, we determined that 25 3CR sources out of 298 were optically unidentified, that is, lacking an optical counterpart of their core, and therefore have neither optical classification nor redshift \citep[see][]{2013ApJS..206....7M}.
{ This means that there is no detected emission from the host galaxy in the optical band, and this could be due to multiple reasons. These sources might, in fact, be either high redshift quasars/radiogalaxies, or highly-absorbed/obscured lower-z active galaxies, or optically faint LERGs, that lack radiatively efficient AGN signatures in the optical emission. 
(see Sect.~\ref{sec:results} for more information on individual sources properties). }

This warranted follow-up observations. 

\citet{2016MNRAS.460.3829M} carried out an optical-to-X-ray campaign, that includes data from the \swf\ Observatory. These authors found that a total of 21 out of the 25 unidentified sources observed by \swf\ have an National Radio Astronomy Observatory (NRAO) Very Large array (VLA) Sky Survey (NVSS, \citealt{1998AJ....115.1693C}) counterpart. 13 of them also show mid-infrared (IR) emission as detected in the AllWISE (Wide-field Infrared Survey Explorer mission, \citealt{2010AJ....140.1868W}) Source Catalog, and nine out of these 21 have an X-ray counterpart detected in the 0.5-10 keV energy range, above 5$\sigma$ level of confidence. 

In this paper, we present the results of \chn\ follow-up observations for seven of the nine unidentified 3C sources with the \swf\ X-ray counterpart, all observed in 2020. The two remaining sources are expected to be observed in April 2021, according to the \chn\ long term schedule\footnote{https://cxc.harvard.edu/target\_lists/longsched.html}, and their analysis will be presented in a forthcoming paper (Missaglia et al., 2021). 

Here, we also present, for the first time, radio observations available for the selected sample in the historical VLA archive. Both radio and X-ray observations are also compared with data collected with the Panoramic Survey Telescope \& Rapid Response System (Pan-STARRS, \citealt{2016AAS...22732407C}) and Wide-field Infrared Survey Explorer (WISE, \citealt{2010AJ....140.1868W}). 

The paper is organized as follows. A brief overview of the data reduction procedures, both for the radio and X-ray band, are given in \S~\ref{sec:obs} while results on single sources are discussed in \S~\ref{sec:results}. In \S~\ref{sec:summary} we present our summary and conclusions. In Appendix \ref{cont}, we show all the radio maps we obtained from the historical VLA archive\footnote{https://science.nrao.edu/facilities/vla/archive/index}. 

Unless otherwise stated, we adopt cgs units for numerical results and we also assume a flat cosmology with $H_0=69.6$ km s$^{-1}$ Mpc$^{-1}$, $\Omega_{M}=0.286$ and $\Omega_{\Lambda}=0.714$ \citep{2014ApJ...794..135B}. Spectral indices, \(\alpha_X\), are defined by flux density, S$_{\nu}\propto\nu^{-\alpha_{X}}$. WISE magnitudes in the nominal bands at 3.4 (W1), 4.6 (W2), 12 (W3), and 22 (W4) $\mu$m are in the Vega system while Pan-STARRS1 adopts the AB magnitude system \citep{1983ApJ...266..713O}. 

\section{Data reduction and analysis}
\label{sec:obs}
To search for optical and infrared counterparts of our selected targets, we firstly retrieved all radio observations from the historical VLA archive, aiming at detecting their radio cores. After data reduction, we overlaid radio contours on optical, IR and X-ray images. In Table \ref{tab:log} we report: (i) 3C designation, (ii,iii) coordinates in J2000 Equinox, (iv) Galactic absorption as reported in \citet{2005A&A...440..775K}, (v,vi) \chn\ observation ID and date, (vii) flux at 178 MHz retrieved from \citet{1985PASP...97..932S}, (viii-xi) remarks on the sources from this work.

\subsection{Radio archival observations}
\label{sec:radio}
All radio data presented in this paper were retrieved from the historical VLA Archive managed by the NRAO. A summary of all radio observations is presented in Table \ref{tab:radio_obs} where we report: (i) 3C designation, (ii) NRAO observing project identification, (iii) observing band, (vi) spectral windows, (v) telescope configuration in which the observation was performed, (vi) clean beam size, (vii) surface brightness of the source, (viii) observation time on source, (ix) Root Mean Square noise of the clean image and (x) contour levels used in the radio maps.

Calibration and imaging were performed in CASA\footnote{https://casa.nrao.edu/} v5.1.1-5 \citep{2007ASPC..376..127M} adopting manual standard procedures. For each source, whenever possible, we reduced observations in L, C and X radio bands (at 1.5, 6 and 10 GHz nominal frequencies, respectively). For all bands, after inspecting the observation log, we manually flagged antennas with bad data. Then, we performed the calibration adopting the following steps: 1) we provided a flux density value for the amplitude calibrator, 2) we derived corrections for the complex antenna gains, 3) we used the flux calibrator to determine the system response to a source of known flux density and finally 4) we applied the calibrations to our calibrators and our target. Bandpass correction is not necessary given that all observations are performed in single channel mode.
As a last step, we performed self-calibration, changing the weight parameter in every step of the cleaning process to recover all the extended emission, and setting manual boxes.

\subsection{X-ray observations}
\label{sec:xrays}
\textit{Chandra} data reduction and analysis have been carried with the \chn\ Interactive Analysis of Observations (CIAO, v4.11; \citealt{2006SPIE.6270E..1VF}) following the standard procedures and threads\footnote{http://cxc.harvard.edu/ciao/threads/}. We also used the \chn\ Calibration Database v4.8.2, according to the same method adopted in our previous investigations of the 3CR snapshot observations (see e.g., \citealt{2010ApJ...714..589M} and \citealt{2011ApJS..197...24M} for additional information). Only a brief overview of the reduction process is reported below. 

We performed the astrometric registration between radio and X-ray images by aligning the X-ray position of each core with that of the radio images, as in previous papers on the 3CR \textit{Chandra} Snapshot Survey (see \citealt{2011ApJS..197...24M};\citealt{2012ApJS..203...31M}; \citealt{2018ApJS..234....7M}; \citealt{2018ApJS..235...32S}). 
In Table \ref{tab:optir}, we report the radio/X-ray shift, that for all sources is less than 2\arcsec, corresponding to $\sim90\%$ of the Chandra Point-Spread Function (PSF).
3CR\,409 is the only source out of the seven presented in this work that could not be registered due to the lack of a radio core detection. WISE and Pan-STARRS datasets are not registered to the radio position, as done for the X-ray data, thus small shifts (i.e., less than $\sim1\arcsec$) may be seen when overlaying radio contours on IR and optical images, consistent with their astrometric uncertainty\footnote{https://wise2.ipac.caltech.edu/docs/release/allwise/;https://panstarrs.stsci.edu
} and seeing in the Pan-STARRS case \citep{2020ApJS..251....6M}. However this does not affect associations of radio and X-ray nuclei with their mid-IR and optical counterparts since below $\sim1\arcsec$ the chance probability of a spurious association is less than 0.1\% \citep[see e.g.,][for details]{2014AJ....148...66M,2019ApJS..242....4D}.

\subsubsection{X-ray photometry}
\label{subsec:photom}
We used unbinned and unsmoothed X-ray images restricted to the 0.5 - 7 keV band to search for the X-ray nuclei. X-ray detection significance, reported as Gaussian equivalent standard deviation ($\sigma$), was estimated measuring the number of photons in both the nuclear region, if present, and a background region. The background region was chosen to be a circular aperture with a radius of 10\arcsec, i.e. 5 times larger in radius than the one used for the X-ray detection of nuclei, and located far enough from the radio galaxy (i.e., at least a few tens of arcsec) to avoid the off-axis degradation of the PSF on Charge Coupled Device borders and contamination from the source, if extended. Adopting a Poisson distribution for the number of photons in the background, we computed the level of significance for X-ray excesses associated with the position of radio cores, if any. For 3CR\,409, where no registration was possible, we centered the nuclear regions at the position corresponding to the emission peak in the 4 - 7 keV band.

We also created flux maps in the X-ray energy ranges: 0.5 -- 1 keV (soft), 1 -- 2 keV (medium), 2 -- 7 keV (hard), taking into account both exposure maps and effective areas. To this end, we used monochromatic exposure maps set to the nominal energies of 0.75, 1.4, and 4 keV for the soft, medium and hard band, respectively. All flux maps were converted from units of $\text{counts}\,\text{cm}^{-2}\,\text{s}^{-1}$ to cgs units by multiplying each map pixel by the nominal energy of each band. We made the necessary correction to recover the observed $\text{erg}\,\text{cm}^{-2}\,\text{s}^{-1}$, when performing X-ray photometry \citep[see e.g.,][for details]{2018ApJS..238...31M,2012MNRAS.424.1774H}. This is the same procedure adopted for X-ray photometry in all previous analyses of our 3CR \chn\ Snapshot Survey \citep[see e.g.,][]{2015ApJS..220....5M,2018ApJS..235...32S,2020ApJS..250....7J} . 

Flux maps were then used to measure observed fluxes for all the X-ray detected nuclei and extended components associated with radio structures. Uncertainties are computed assuming Poisson statistics (i.e., square root of the number-of-counts) in the source and background regions. X-ray fluxes for the cores are not corrected for the Galactic absorption, and are reported in Table~\ref{tab:cores}.

\subsubsection{X-ray surface brightness profiles}
\label{subsec:surbri}

For 3CR\,409 and 3CR\,454.2, we detected significant diffuse X-ray emission in the 0.5 - 7 keV band, extending beyond the radio structure. Thus, to estimate the extension of this X-ray emission, we derived its surface brightness profiles, reported in Sec. \ref{sec:results}).

Firstly, we detected and removed X-ray point-like sources (including the X-ray nuclei of the radio galaxies) in the 0.5 - 7 keV images using the \textsc{wavdetect} task, available in CIAO. We adopted a sequence of $\sqrt{2}$ wavelet scales, from 1 to 16 to cover different sized sources, and a false-positive probability threshold set to the value of $10^{-6}$, which is the value recommended for a 1024 $\times$ 1024 image in the CIAO threads\footnote{https://cxc.harvard.edu/ciao/threads/wavdetect/}. This value was chosen to ensure that we do not over-subtract point sources. We generated the corresponding source regions using the \textsc{roi} task. 

We computed the 0.5 - 7 keV, exposure corrected X-ray surface brightness profiles in four quadrants (north, south, east, west). 
The background was estimated as a circular region of $\sim$ 80\arcsec\ radius, far from the source, and free of detected sources. A similar procedure was used in \citet{2021ApJS..252...31J} to search for X-ray counterparts of radio hotspots.

\subsubsection{X-ray spectral analysis}
\label{subsec:spec}
We performed spectral analysis for the X-ray counterparts of radio cores of four sources having more than 400 photons (as reported in Table \ref{tab:cores}) within a circular region of 2\arcsec, centered on the radio core position (namely, 3CR\,91, 3CR\,390, 3CR\,409 and 3CR\,428) and in more extended regions corresponding to X-ray diffuse emission. This analysis was carried out to determine X-ray spectral indices \(\alpha_X\), the presence of significant intrinsic absorption, if any, and to estimate the temperature, abundances and density for the intracluster medium (ICM) of the two galaxy clusters detected (namely 3CR\,409 and 3CR\,454.2). 

The spectral data for the X-ray cores were extracted from a 2\arcsec\ aperture, as for photometric measurements, using the CIAO routine \textsc{specextract}, thereby automating the creation of count-weighted response matrices. { In the cases of the sources hosted in clusters, namely 3CR\,409 and 3CR\,454.2, we have extracted the spectrum from a circular region of $\sim$ 1' excluding a 2\arcsec\ aperture covering the nucleus.} Background spectra were extracted in nearby circular regions of radius 80\arcsec\ free of detected sources. The source spectra were then filtered in energy between 0.5-7 keV and binned to allow a minimum number of { 20} counts per bin to ensure the use of the $\chi^2$ statistic. We used the \textsc{Sherpa}\footnote{\href{http://cxc.harvard.edu/sherpa}{http://cxc.harvard.edu/sherpa}}\citep{2001SPIE.4477...76F} modeling and fitting package to fit our spectra. 

For the nuclear spectra, as performed in all our previous analyses of sources observed during the 3CR \chn\ Snapshot Survey, { we adopted an absorbed power-law model with the hydrogen column density $N_H$ fixed at the Galactic values and a contribution of intrinsic absorption (xswabs*xszwabs*xspowerlaw)}. In the following section, we report the values obtained by our best fits. When considering the fitting model, the two main free parameters - namely the intrinsic absorption $N_{H,int}$ and the spectral index \(\alpha_X\) - were allowed to vary, to quantify the degree to which $N_{H,int}$ and \(\alpha_X\) are degenerate. In all cases, we adopted the { photometric redshift} obtained from WISE counterpart magnitudes, as described in \citet{2017arXiv170908634G}. { These authors, using the WISE two-color plot (W1-W2 versus W2-W3), were able to distinguish LERGs (typically associated with passive elliptical galaxies), HERGs (associated with smaller but higher star-forming galaxies hosting radiative AGNs) and QSOs. The method described to estimate the redshift offers, therefore, also the probability of a radio source to be either a LERG, or a HERG or a QSO, using the kernel density estimation (KDE) (for more information about this method see Section 2.2 in \citealt{2017arXiv170908634G}). These probabilities are used to weight the redshift estimation made for each class (see Table 2 in \citealt{2017arXiv170908634G}). In Table \ref{tab:xrayspec} photometric redshifts and results of the X-ray spectral analysis of the cores are reported.}

For 3CR\,91, from the pileup map generated through the CIAO task pileup\_map, we estimated the amount of pileup to be $\sim$ 15\%. For this reason, before performing the spectral analysis of the X-ray counterpart of the radio core, we excluded the pixels most affected by the pileup (five pixels), obtaining a better fit with respect to the one adopting the jdpileup \footnote{https://cxc.cfa.harvard.edu/sherpa/ahelp/jdpileup.html} model.

\section{Results and details on individual sources}
\label{sec:results}

\begin{figure}[ht!]
\centering
\includegraphics[scale=0.17]{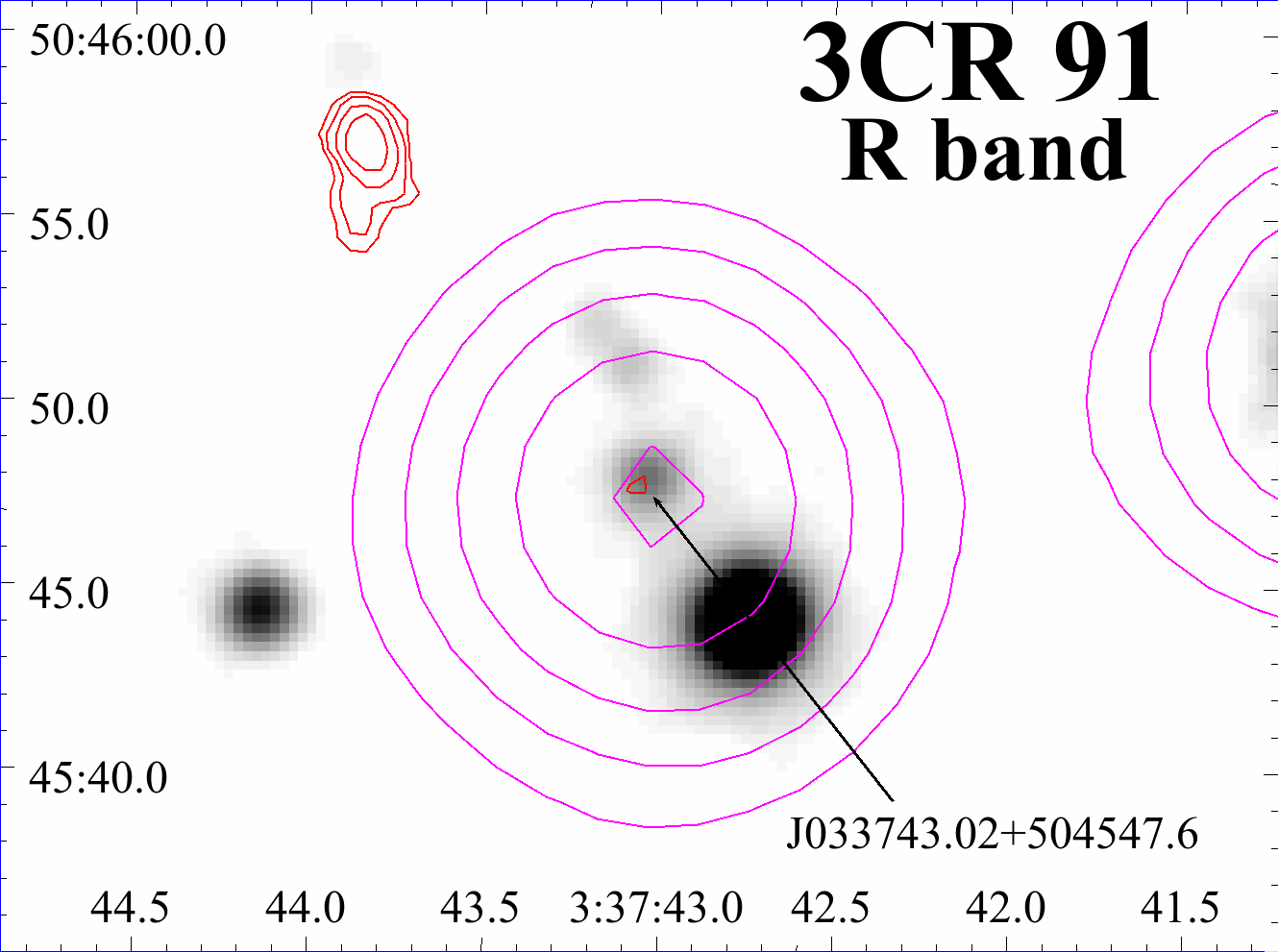}
\includegraphics[scale=0.17]{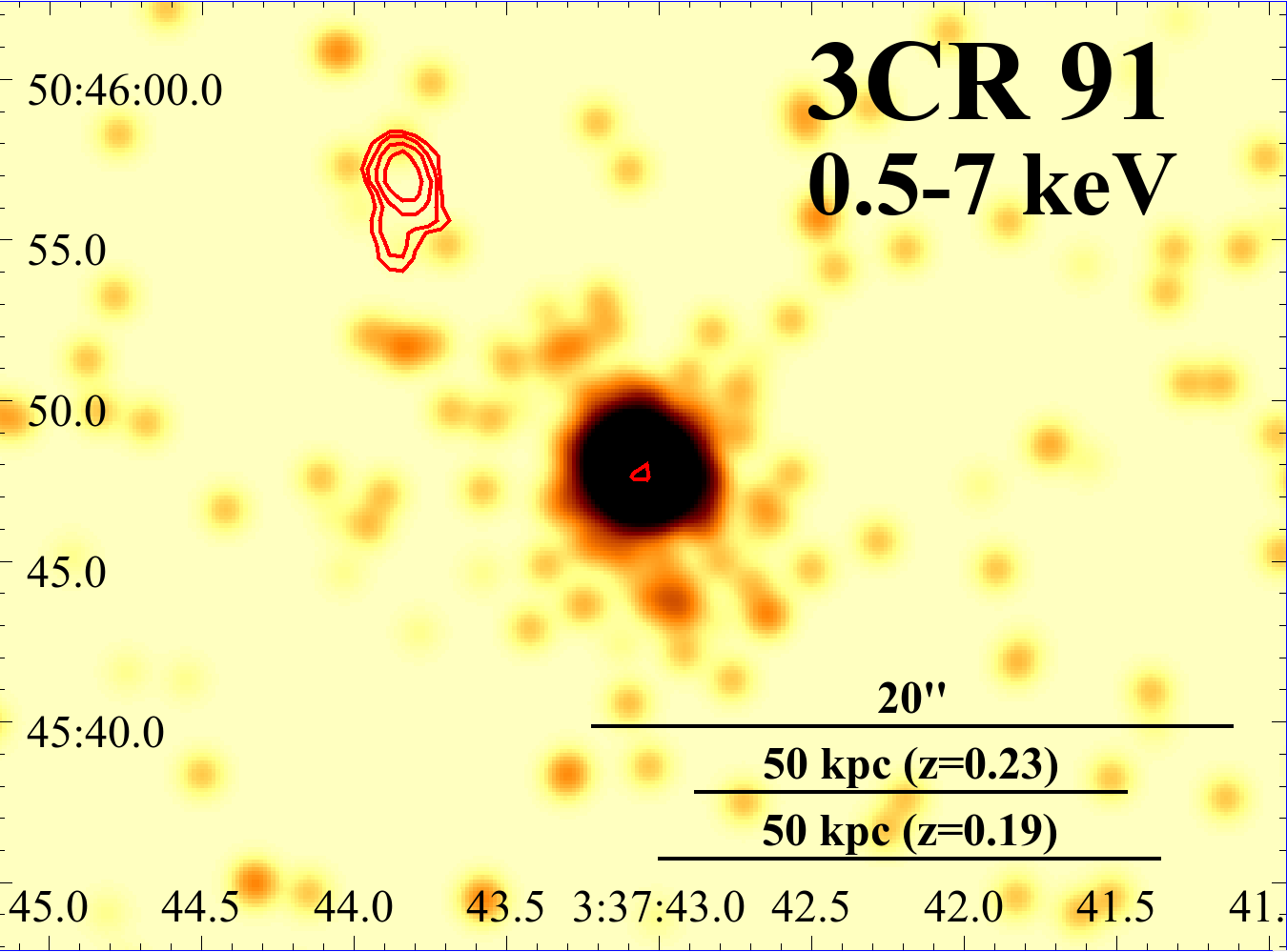}
\caption{(Upper Panel): Pan-STARRS R band image with WISE 3.4~$\mu$m filter magenta contours of the IR counterpart of the radio source overlaid. IR contours are drawn at 12.54, 16.91, 22.89, 31.07 and 42.27 in arbitrary flux scale. In red, 8 GHz VLA contours are shown, the same used in the \chn\ image. The four radio contour levels were computed starting at 0.01 Jy/beam, increasing by a factor of 2. The arrow points to the WISE counterpart of the radio core. (Lower panel): 0.5-7 keV \chn\ image of 3CR\,91 with VLA 8 GHz contours overlaid. The image has not been rebinned, but smoothed with a 3 pixel (equivalent to 1.48\arcsec) Gaussian kernel. { In the bottom right of the image, kpcscale measured using the photometric redshifts obtained using the method described in \citet{2017arXiv170908634G} are indicated. }}
\label{fig:3C91chn}
\end{figure}

\subsection{3CR\,91}
From the historical VLA archive, we retrieved radio observations of 3CR\, 91 performed at 1.4 and 8\,GHz in B and AB configurations, respectively. In the 8\,GHz image (see red contours in the panels of Fig.~\ref{fig:3C91chn} and right panel of Fig.~\ref{3c91_radio}) the radio core is clearly detected (i.e., above 5$\sigma$ confidence level), while in the VLA image at 1.4\,GHz (see Fig.~\ref{3c91_radio}, left panel) the radio core is not visible.
3CR\,91 is a double-lobed radio source in the VLA image at 1.4\,GHz.
On the other hand, in the 8\,GHz radio map, we did not detect the emission arising from the southern radio lobe, clearly seen at lower frequencies, { but the northern lobe resembles an FRII radio galaxy}.

We found both the IR and the optical counterpart of the radio core in the Pan-STARRS and WISE images, as shown in the upper panel of Fig.~\ref{fig:3C91chn}. The WISE counterpart to the core, J033743.02+504547.6, detected in all IR filters, is clearly the same object found within the \swf\ XRT uncertainty circle, at small angular separation (1.4\arcsec) from its radio position. This source, as reported in \citet{2016MNRAS.460.3829M}, has an associated NVSS counterpart, J033743+504552. 
The WISE counterpart is included in \citet{2014ApJS..215...14D} all-sky catalogue of infrared selected, radio-loud active galaxies, due to its peculiar infrared colors, and also in \citet{2019ApJS..242....4D}. Since 3CR\,91 has a WISE counterpart, adopting the method described in \citet{2017arXiv170908634G}, we were able to obtain a { photometric redshift estimate of $z=0.23\pm0.18$ from the 3.4~$\mu$m magnitude, and an estimate of $z=0.19^{+0.18}_{-0.14}$ from the 4.6~$\mu$m magnitude, with 82\% probability of being a QSO}. 
In the \chn\ image (Fig.~\ref{fig:3C91chn}, lower panel), there is a clear detection of the radio core in the 0.5-7 keV energy range. 3CR\,91 also shows extended X-ray emission up to $\sim$9\arcsec\ from the nucleus. We did not detect any X-ray counterpart for hotspots and lobes. 

Since for 3CR\,91 the number of photons within a circular region of 2\arcsec\ radius, centered on the radio position, is larger than 400, we carried out a nuclear X-ray spectral analysis. { We adopted an absorbed power-law model with the hydrogen column density $N_H$ fixed at the Galactic value (see Table \ref{tab:log}, (iv) column). As reported in Table~\ref{tab:xrayspec}, for both photometric redshifts, we obtained the best fit values setting $N_{H,int}$ as free parameter, obtaining a value \(\sim {10}^{23} \text{ cm}^{-2}\), similar to that of the Galactic $N_H$. Since 3CR\, 91 is a moderate-z QSO, we did expect a detection of the optical counterpart, in agreement with our results.}

\subsection{3CR\,131} 
\begin{figure}[ht!]
\centering
\includegraphics[scale=0.2]{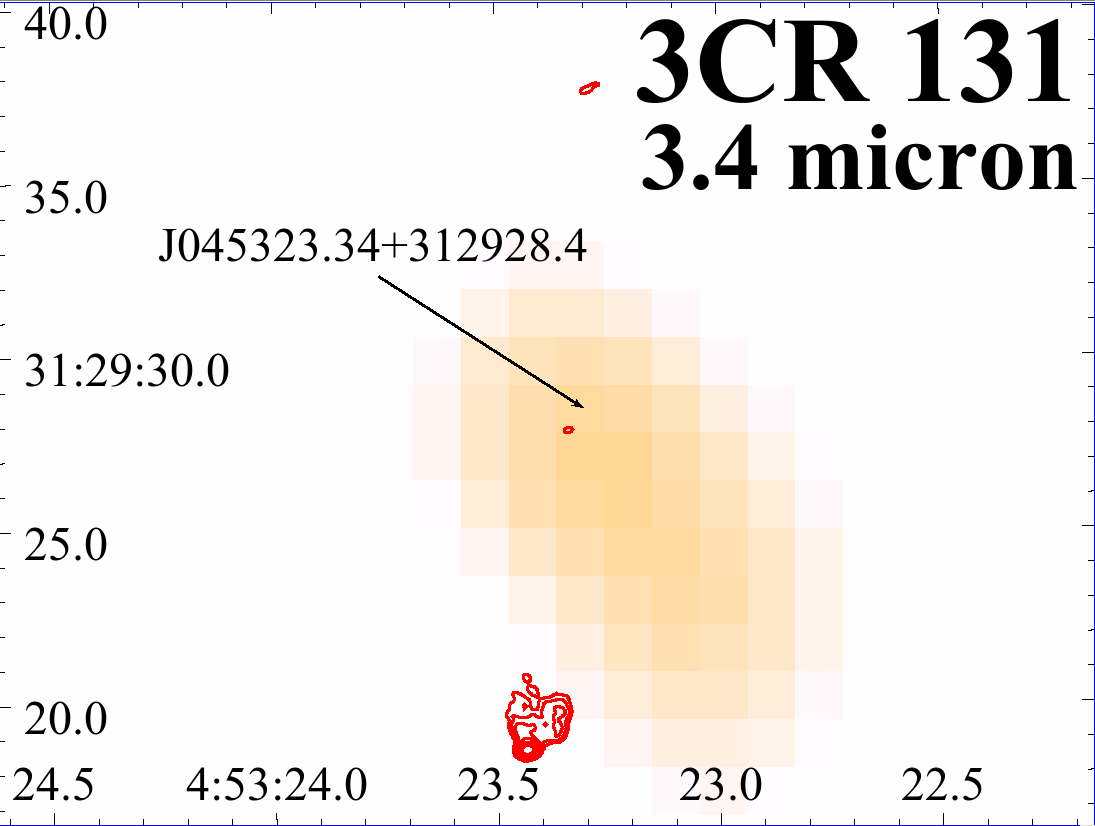}
\includegraphics[scale=0.21]{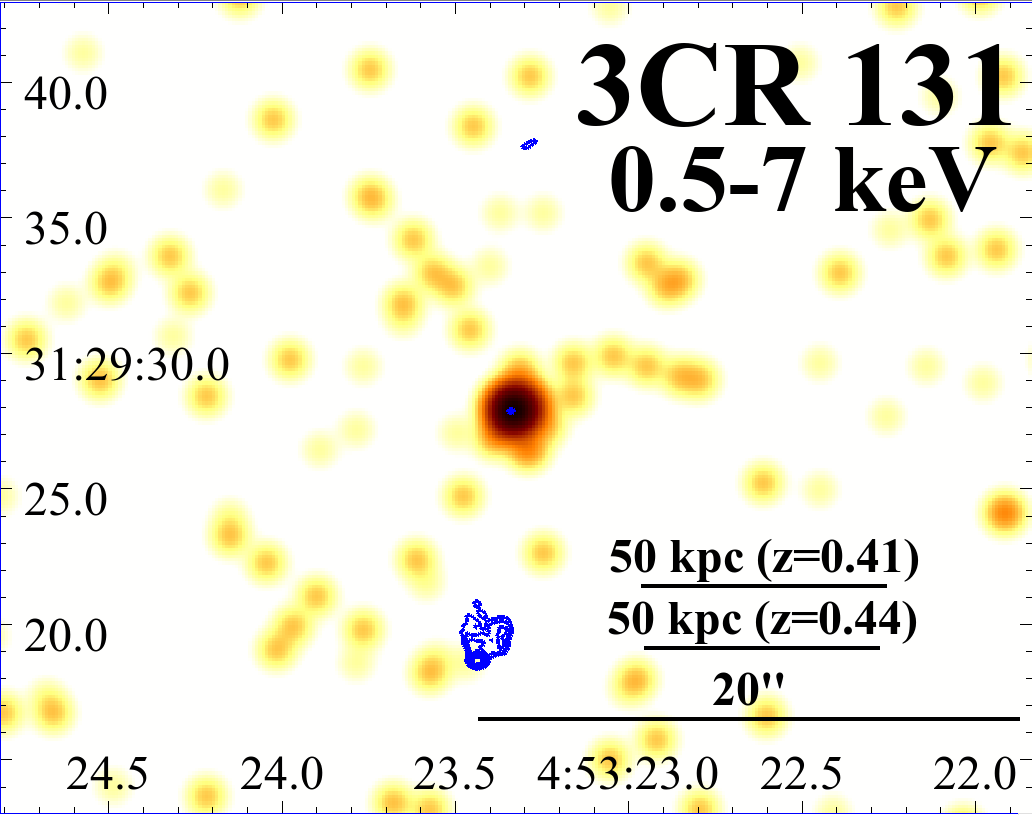}
\caption{(Upper panel): WISE 3.4~$\mu$m filter image of 3CR\,131. The arrow points to the IR counterpart of the radio nucleus. No optical counterpart has been detected for this source. The six VLA 8\,GHz radio contours levels (red) starts from 0.002 Jy/beam and are increased by a factor of 2. (Lower panel): 0.5-7 keV \chn\ image of 3CR\,131, with VLA 8 GHz band contours overlaid in red. \chn\ image is not rebinned, but has been smoothed with a 5 pixel (2.46\arcsec) Gaussian kernel. { In the bottom right of the image, kpcscale measured using the photometric redshifts obtained using the method described in \citet{2017arXiv170908634G} are indicated. } }
\label{fig:3C131chn}
\end{figure}

For 3CR\,131, we reduced VLA data at 8\,GHz (see red radio contours in Fig.~\ref{fig:3C131chn}, both panels) and at 1.4\,GHz (see Fig. \ref{3c131_radio} right panel). In the 8\,GHz image, we detected the emission of both lobes and the nucleus, at 5 times the rms noise level, and a single radio hotspot in the southern lobe. { The presence of the northern hotspot could suggest that 3CR\,131 is an FRII radio galaxy. Using the method described in \citet{2017arXiv170908634G}, we obtained a photometric redshift estimate of $z=0.41^{+0.13}_{-0.12}$ from the 3.4~$\mu$m magnitude value, and an estimate of $z=0.4\pm0.13$ the 4.6~$\mu$m magnitude, and a 59\% probability for this source to be a LERG. }

We found the nuclear counterpart of 3CR\,131 only in the IR image. In WISE $W_{1}$ filter (3.4~$\mu$m) image (see Fig.~\ref{fig:3C131chn}, upper panel), there are two nearby objects at angular separation $4\arcsec$ $\sim$ from the position of  NVSS  J045323+312924 \citep{2016MNRAS.460.3829M}. Only one object, WISE J045323.34+312928.4 (indicated by an arrow in Fig.~\ref{fig:3C131chn} upper panel), is within the positional uncertainty of the \swf-XRT source. This WISE source is cospatial with the position of the nucleus in the 8\,GHz radio map, and it is therefore likely its IR counterpart.

In the \chn\ image (Fig.~\ref{fig:3C131chn}, lower panel) the core is clearly detected and associated with the radio core, but the southern hotspot is not detected in the 0.5 - 7 keV energy range.

\subsection{3CR\,158} 
\begin{figure}[ht!]
\centering
\includegraphics[scale=0.22]{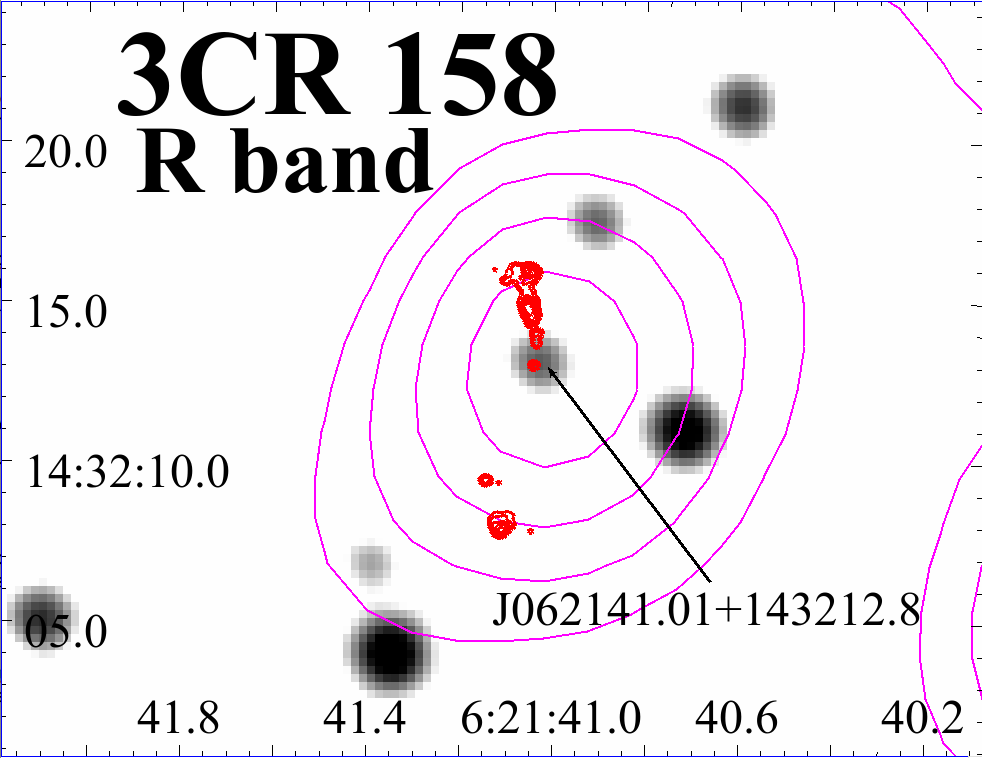}
\includegraphics[scale=0.21]{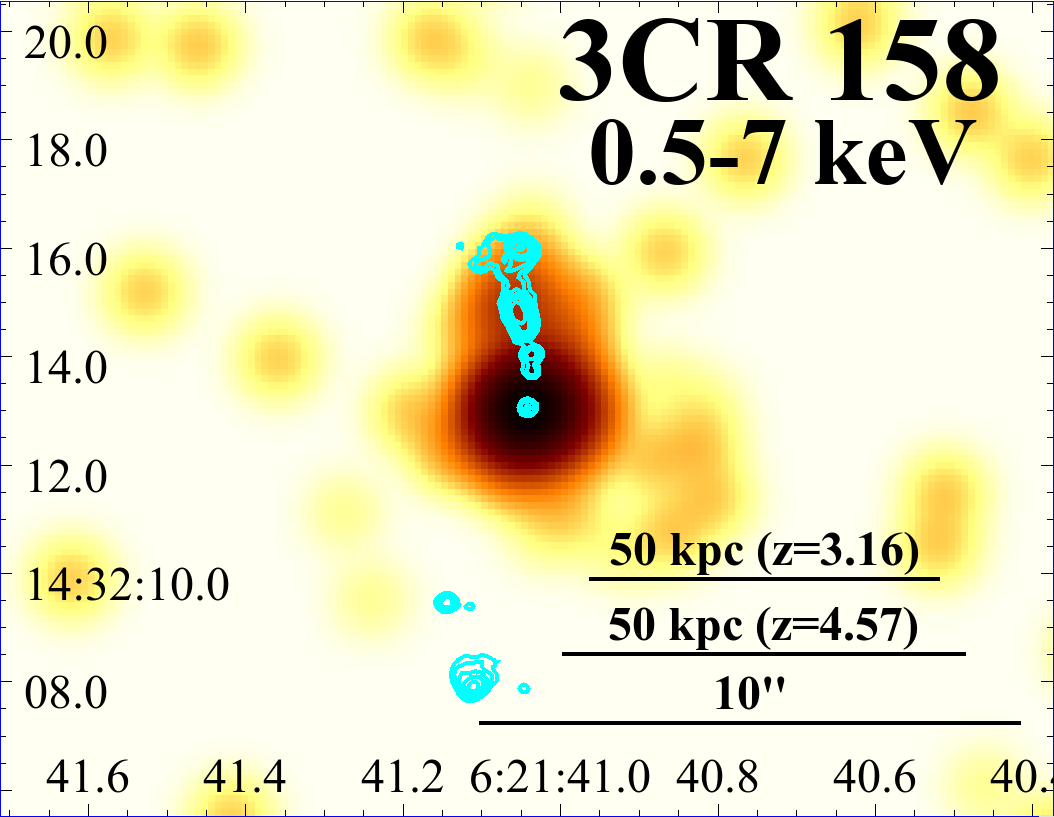}
\caption{(Upper panel): Pan-STARRS R band image with WISE 3.4~$\mu$m filter magenta contours of the IR counterpart of the radio source overlaid. In red 8\,GHz VLA contours are shown, the same used in the \chn\ image (in cyan). VLA contours start from 0.002 Jy/beam, increasing by a factor of 2 up to 0.064 Jy/beam. IR contours are drawn at 13.40, 13.75, 14.21, 14.80 in arbitrary flux scale. The arrow points to the WISE counterpart of the radio core. (Lower panel): \chn\ X-ray image of 3CR\, 158 in the 0.5-7 keV energy band, binned to 0.123\arcsec/pixel and smoothed with a 5 pixel Gaussian kernel (equivalent to 0.615\arcsec).  The northern radio jet has a clear X-ray counterpart. { In the bottom right of the image, kpcscale measured using the photometric redshifts obtained using the method described in \citet{2017arXiv170908634G} are indicated. }}
\label{fig:3c158chn}
\end{figure}

3CR\,158, at 8\,GHz, is a lobe dominated radio source (see red/cyan contours in Fig.~\ref{fig:3c158chn}). The core is clearly detected in this band, as well as both the southern and northern lobes (see red and cyan contours in both panels of Fig.~\ref{fig:3c158chn} and upper left panel of Fig.~\ref{3c158_radio}). In particular, in the northern side there are two knots and one hotspot, { hint that the source could be classified as FRII}. The two knots in the southern radio structure are probably part of the same lobe but are not detected in the 4.5\,GHz image (see Fig.~\ref{3c158_radio} upper right and bottom panels). 

The core is also detected in both optical and IR images (see Fig.~\ref{fig:3c158chn} bottom panel). \citet{2016MNRAS.460.3829M} reported that the X-ray source XRT J062141.2+143212 matches the NVSS source J062141+143211, within the 3C positional uncertainty region of 3CR\,158. An infrared counterpart in the AllWISE Source Catalogue, WISE J062141.01+143212.8, is found at 1.5\arcsec\ from the NVSS source, probably its counterpart (we remind that at these separations the chance probability of spurious association is below 0.1\%). No infrared/optical candidate counterpart has been previously reported in the literature for this radio source. { Adopting the procedure described in \citet{2017arXiv170908634G}, using the 3.4~$\mu$m WISE magnitude we obtained a photometric median redshift of $z=4.57^{+0.68}_{-3.14}$ and using the 4.6~$\mu$m WISE magnitude, a photometric median redshift of $z=3.16^{+2.09}_{-2.19}$, with a probability of 82\% for this source of being a QSO. This is the source with the highest predicted photometric redshift of the sample. Since the Pan-STARRS counterpart is detected, even at such high redshift, a spectroscopic optical campaign is required to verify this prediction.}

We found X-ray extended emission aligned with the radio jet structure in the northern side of our \chn\ observation (see Fig.~\ref{fig:3c158chn} upper panel). The flux of of the X-ray counterpart is (1.51$\pm$0.30)$\times$10$^{-14}\,\text{erg}\,\text{cm}^{-2}\,\text{s}^{-1}$ in the 0.5-7 keV band. Given the alignment with the jet rather than the lobe, the jet is most likely the source of the X-rays.

\subsection{3CR\,390} 
\begin{figure}[ht!]
\centering
\includegraphics[scale=0.18]{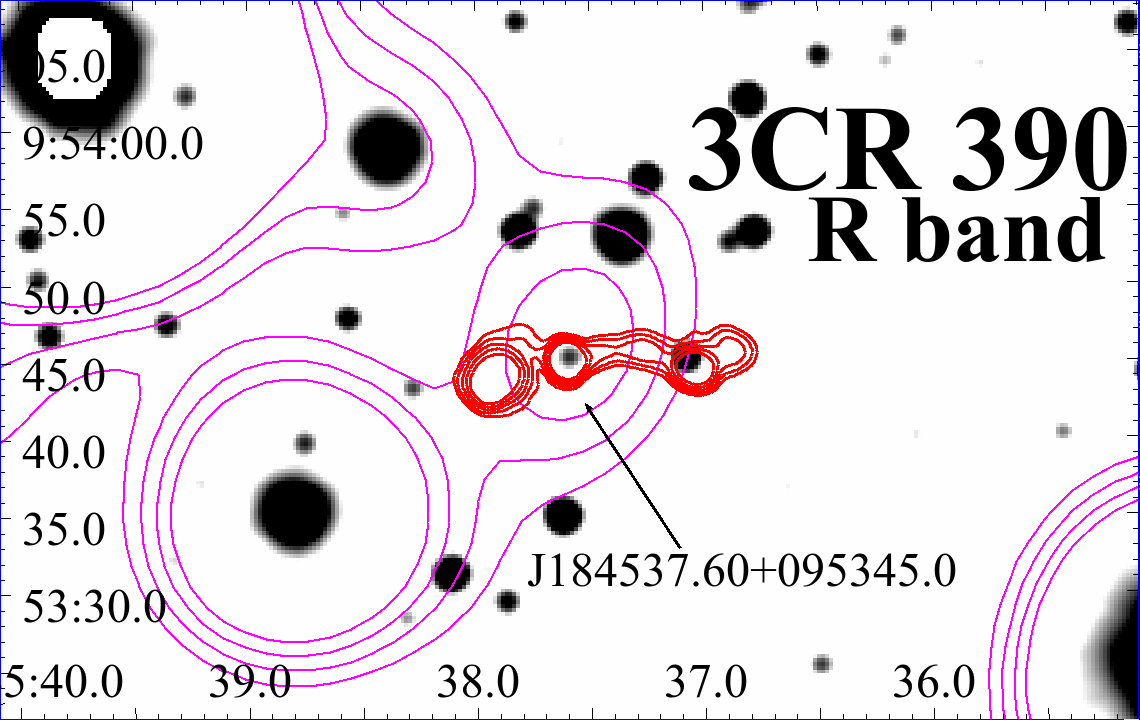}
\includegraphics[scale=0.18]{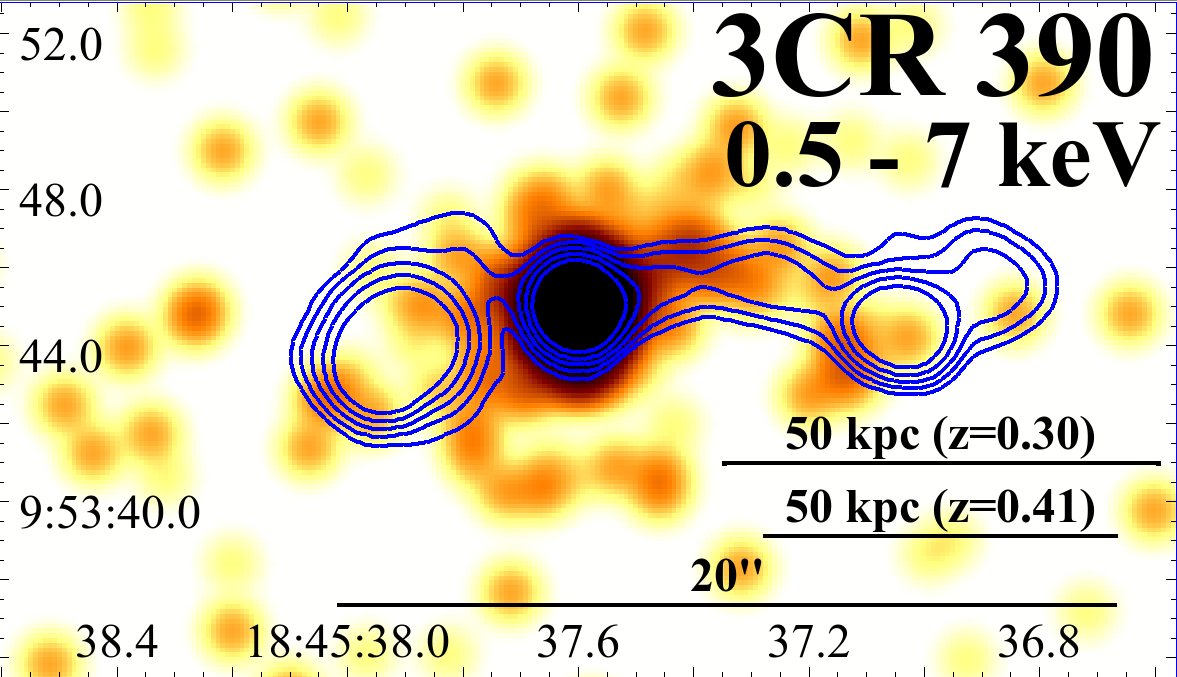}
\caption{(Upper panel): Pan-STARRS R band image of the field of 3CR\,390, with WISE 3.4~$\mu$m filter magenta contours overlaid. IR contours are drawn at 16.64, 19.28, 22.42, 26.15 in arbitrary flux scale. VLA red contours at 4.5\,GHz are overlaid, starting from 0.002 Jy/beam  and increased by a factor of two up to 0.032 Jy/beam. The black arrow points to the position of the IR counterpart of the radio nucleus. (Lower panel): \chn\ X-ray image of 3CR\,390 in the 0.5-7 keV band, binned to 0.246\arcsec/pixel and smoothed with a 5 pixel Gaussian kernel (equivalent to 1.23\arcsec). The X-ray image shows extended emission spatially coincident with the radio bridge in the western direction. { In the bottom right of the image, kpcscale measured using the photometric redshifts obtained using the method described in \citet{2017arXiv170908634G} are indicated. }}
\label{fig:3c390chn}
\end{figure}

3CR\,390 is a lobe dominated radio source, as shown in the VLA image at 4.5\,GHz (see red/blue contours in both panels of Fig.~\ref{fig:3c390chn} and Fig.~\ref{3c390_radio} upper left panel). The IR counterpart of the radio core is detected in the WISE 3.4~$\mu$m filter image (see Fig.~\ref{fig:3c390chn}, upper panel). 

In the Pan-STARRS R band image, there are both an optical source corresponding to the radio core, in the same position of the WISE counterpart, and an optical source located on the position of the western radio knot (see Fig.~\ref{fig:3c390chn}, lower panel) that is likely to be a background source. 
To verify if the X-ray emission on the western side of the source is related to the diffuse radio emission or if it is an optical source, we measured the p-chance of the hotspot association considering the source density in a 80\arcsec\ circular region around 3CR\,390, taking into account the correct distance between the core and tentative hotspot. We obtained a p-chance $<$ 4\%. As reported in \citet{2016MNRAS.460.3829M}, the whole radio structure is associated with NVSS J184537+095344 and the X-ray source: XRT J184537.6+095349. The NVSS source, not well centered with respect to the 3CR positional uncertainty region, is located within the XRT positional uncertainty region. The infrared counterpart, namely: WISE J184537.60+095345.0, was also included in the all-sky catalogue of blazar candidates by \citet{2014ApJS..215...14D} due to its peculiar infrared colours, and in \citet{2019ApJS..242....4D}.
{ Adopting the procedure described in \citet{2017arXiv170908634G}, using the 3.4 $\mu$m WISE magnitude, we obtained a photometric median redshift of $z=0.41^{+0.28}_{-0.32}$, and using the 4.6 $\mu$m WISE magnitude a photometric median redshift of $z=0.30^{+0.21}_{-0.25}$, with a probability of  57\% for the source to be a QSO.}

In the \chn\ image the core is clearly detected and we also found extended X-ray emission spatially coincident with the radio bridge connecting the two intensity peaks visible at 4.5\,GHz (see Fig.~\ref{fig:3c390chn} lower panel). Given the large number of X-ray photons measured within 2\arcsec\ circular region from the radio core position (i.e., above our threshold of 400 counts) we performed the X-ray spectral analysis. 
Adopting a power-law model { with both Galactic and intrinsic absorption, we obtained the best fit results, for both values of the photometric redshifts, setting $N_{H,int}$ as free parameter. In these cases, we obtain an $N_{H,int}$ that is comparable with the Galactic $N_H$ (\(\sim {10}^{22} \text{ cm}^{-2}\)), and this result is not unexpected given that we were able to detect the optical counterpart in Pan-STARRS and the moderate-z of 3CR\,390.}


\subsection{3CR\,409} 
\begin{figure}[ht!]
\centering
\includegraphics[scale=0.2]{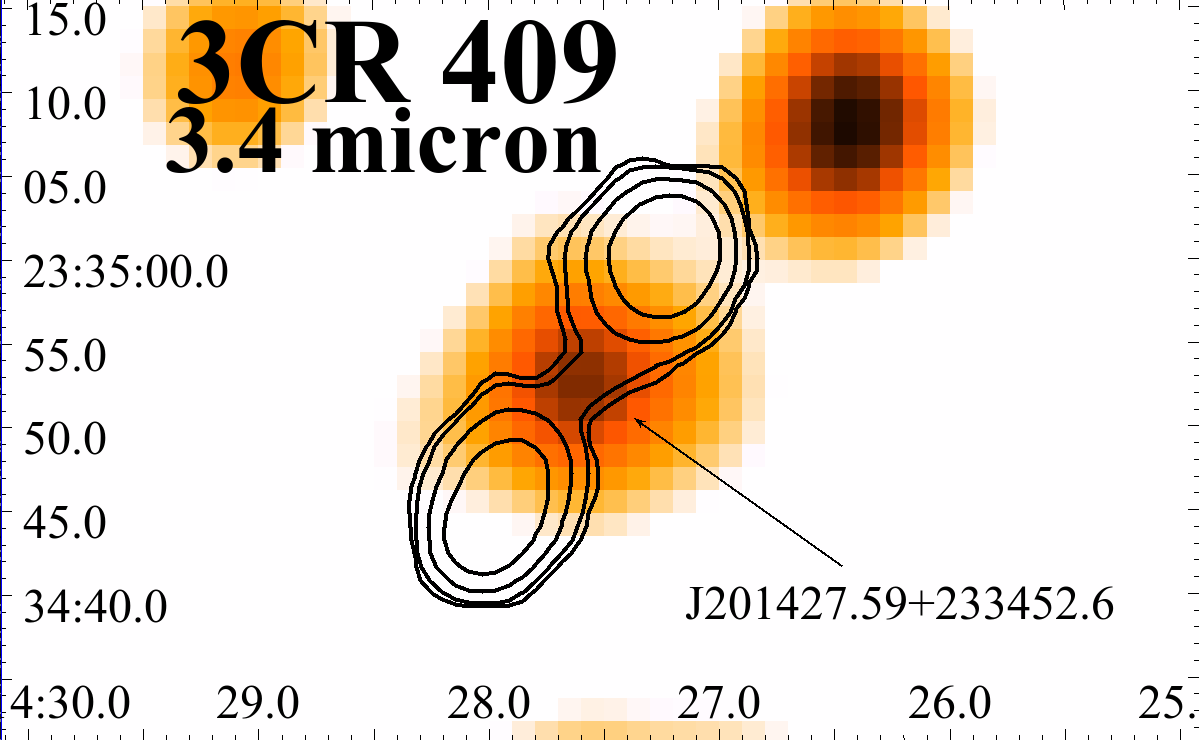}
\includegraphics[scale=0.18]{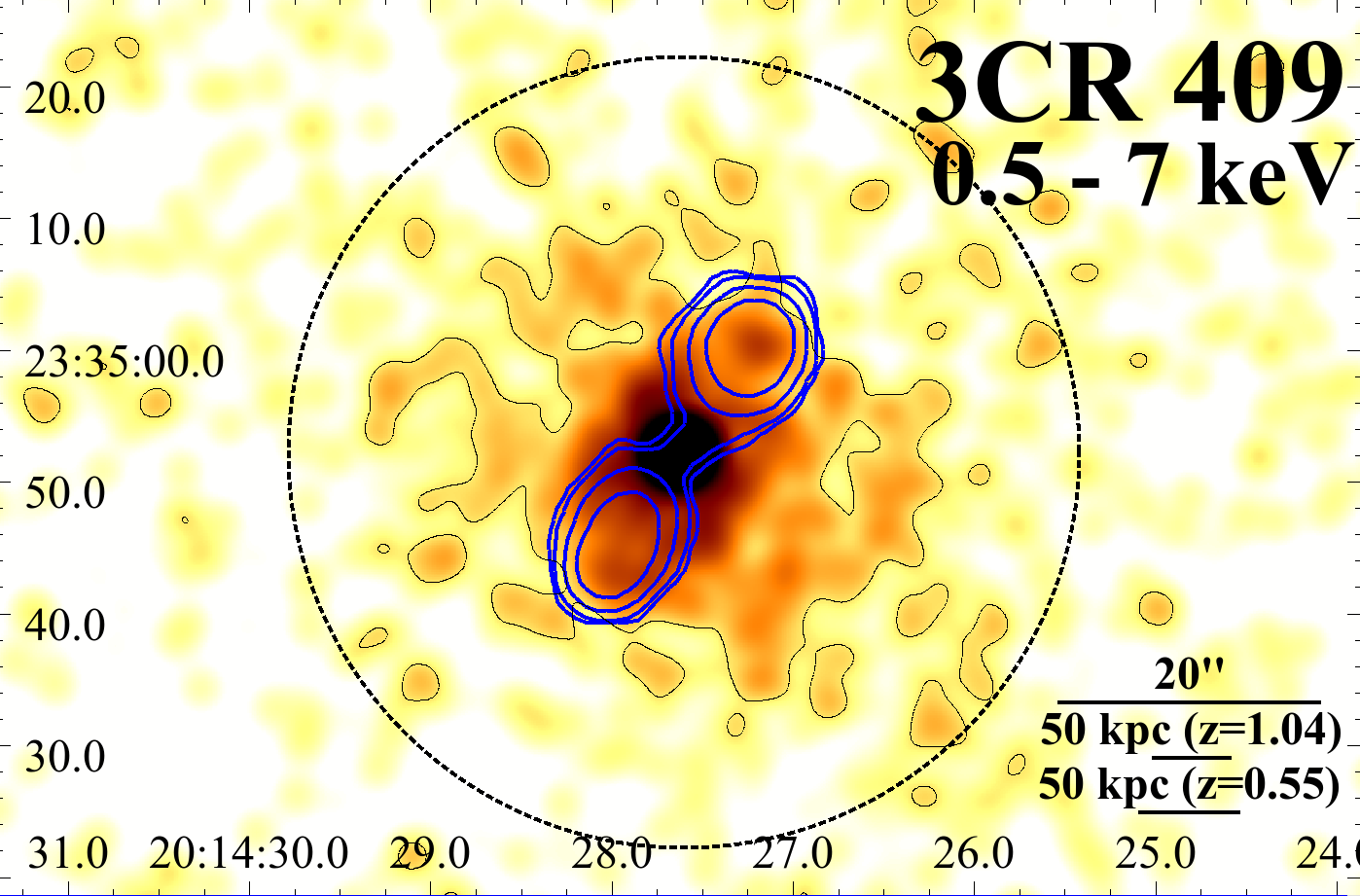}
\caption{(Upper panel): WISE 3.4~$\mu$m filter image of 3CR\,409. The arrow points to the IR counterpart of the radio nucleus. No optical counterpart has been detected for this source. VLA red contours at 4.5\,GHz are overlaid, starting from 0.02 Jy/beam increasing by a factor of two, up to 0.64 Jy/beam. (Lower panel): \chn\ X-ray image of 3CR\,409, filtered in the 0.5-7 keV band, rebinned to 0.123 \arcsec/pixel and smoothed with a 4.92\arcsec Gaussian kernel. VLA contours are the same used in the upper panel. The dotted black circle has a 0.5\arcmin\ radius. The black contours trace the X-ray emission at $3\times10^{-18}\,\text{erg}\,\text{cm}^{-2}\,\text{s}^{-1}$. The X-ray image shows emission associated with the radio lobes as well as more extended emission extending to $60\arcsec$. { In the bottom right of the image, kpcscale measured using the photometric redshifts obtained using the method described in \citet{2017arXiv170908634G} are indicated. } }
\label{fig:3c409chn}
\end{figure}

For 3CR\,409, a lobe dominated radio source, we merged two VLA observations, both at 1.4\,GHz, obtained in different configurations (see column (v) of Table \ref{tab:radio_obs} and black/blue contours in Fig.~\ref{fig:3c409chn}). { This is the only radio source of our sample to have been previously classified, as an FRII \citep{2013ApJS..206....7M}.} We manually flagged and calibrated both datasets separately, and then we performed the self-calibration of both observations together. In the final radio map (see Fig.~\ref{3c409}) the core is not clearly detected, and overlaying radio contours to the optical image we did not find a plausible counterpart located between the two lobes. On the other hand there is a WISE source, namely: J201427.59+233452.6, detected in all four filters and correspondent to the intensity peak of the \chn\ image. 

The source \swf\ J201427.5+233455, detected with S/N=11.6, lies in the field of view of 3CR\,409 and matches the coordinates of NVSS J201427+233452 with an angular separation of 1.9\arcsec. The mid-IR counterpart WISE J201427.59+233452.6 is located at only 0.3\arcsec from the NVSS source. This WISE infrared object is included in the all-sky catalogue of blazar-like radio-loud sources by \citet{2014ApJS..215...14D} and in \citet{2019ApJS..242....4D}. { As for previous sources, adopting the procedure described in \citet{2017arXiv170908634G}, using the 3.4~$\mu$m WISE magnitude we obtained a median redshift of $z=1.04^{+0.71}_{-0.74}$, and using the 4.6~$\mu$m WISE magnitude a median redshift of $z=0.55^{+0.38}_{-0.46}$ and a probability of 60\% of being a QSO.}

\begin{figure*}
\centering
\includegraphics[scale=0.25]{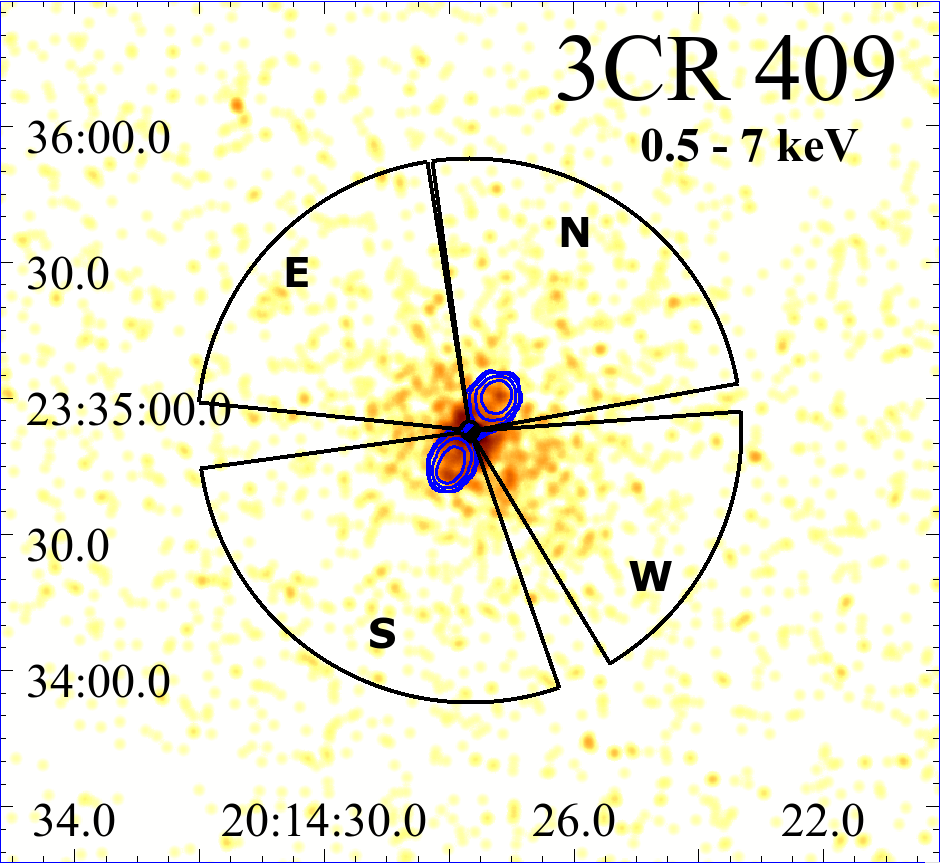}
\caption{Directions used in source 3CR\,409 to extract the surface brightness profiles shown in Fig.~\ref{sur_bri}. The four sectors extend up to $60\arcsec$ from the core (that we have excluded, starting from a distance of $2\arcsec$ from the position of the core, as reported in the NVSS, see Table \ref{tab:log}). In blue we show the VLA contours from Fig.~\ref{fig:3c409chn}.}
\label{409_sur_bri_regions}
\end{figure*}

In the \chn\ image there is some extended emission around the core, suggesting the presence of a cluster, that we investigate in more details. We have derived the surface brightness profiles in the directions shown in Fig.~\ref{409_sur_bri_regions}. Northern and southern directions have been selected to encompass the radio lobes contours, while eastern and western directions are that away from the lobes. From these profiles we have estimated the extension of the diffuse emission, that appears to be symmetrical around the source up to a distance of $\sim$ 60\arcsec\ from the radio core. In the west direction, at a distance of $\sim$ 10\arcsec\ there is a jump in the surface brightness, and the same behaviour can be observed at the same distance in the east direction. 

Since we detected more than 400 photons in the nuclear region of the \chn\ image, we also performed the spectral analysis of the X-ray core, adopting { an absorbed power-law model. We obtained the best fit results with $N_{H,int}$ as a free parameter, for both choices of redshift. $N_{H,int}$ has a value of \(\sim {10}^{23} \text{ cm}^{-2}\), and this is probably the reason we did not detect an optical counterpart, under the assumption of a normal gas to dust ratio.


We also analysed the extended X-ray emission, adopting a thermal \textsc{APEC} model with Galactic absorption. As specified in Subsection~\ref{subsec:spec}, we have excluded a 2\arcsec\ circular region where we expect to find most of the nuclear emission and all the detected point sources. However, we also took into account the contribution of \chn\'s PSF wings extending into the region selected for the spectral extraction, that accounts for about $\sim$2\% of the 0.5-7 keV net counts. We therefore included such a contribution in the model used to fit the extended emission. We have tested four models: abundance fixed to 0.25 solar and redshift $z=1.04$ (or $z=0.55$); redshift $z=1.04$ (or $z=0.55$) and free abundance; abundance fixed to 0.25 solar and free redshift, and both abundance and redshift free to vary during the fit. The $z=0.55$ models yield the best fit statistics, with a poorly constrained temperature \(\gtrsim 11 \text{ keV}\). Such temperatures have been reported for clusters at higher redshifts (e.g., \(z=0.89\), \citealt{2004cgpc.sympE..25J}), and therefore deeper \textit{Chandra} observations are needed to draw firm conclusions on the nature of this extended X-ray emission.}


\begin{figure}[ht!]
\centering
\includegraphics[scale=0.2]{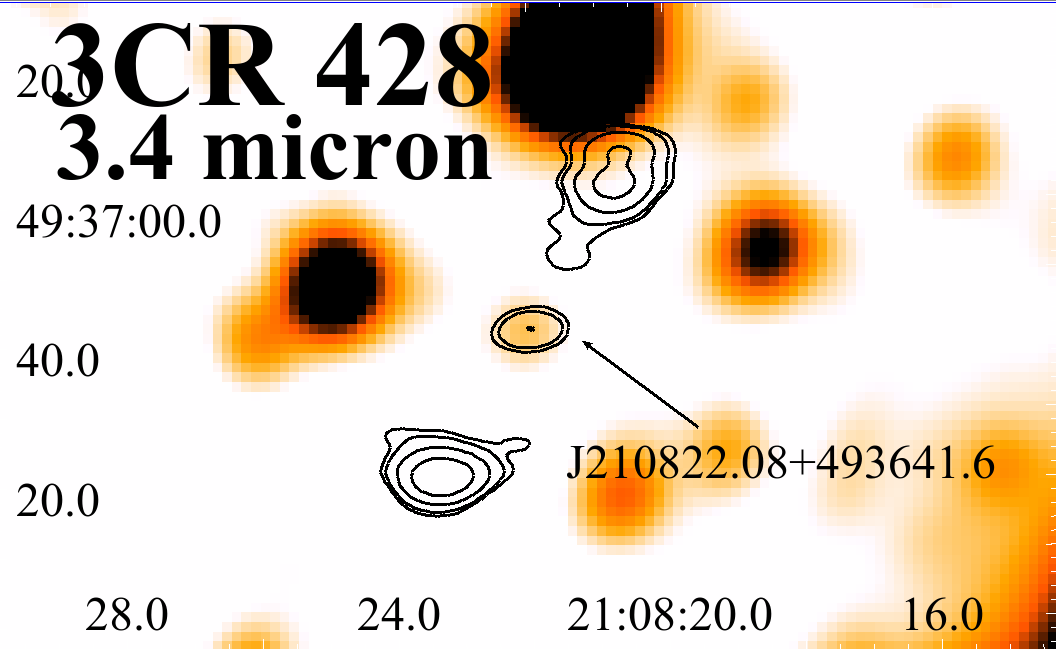}
\includegraphics[scale=0.2]{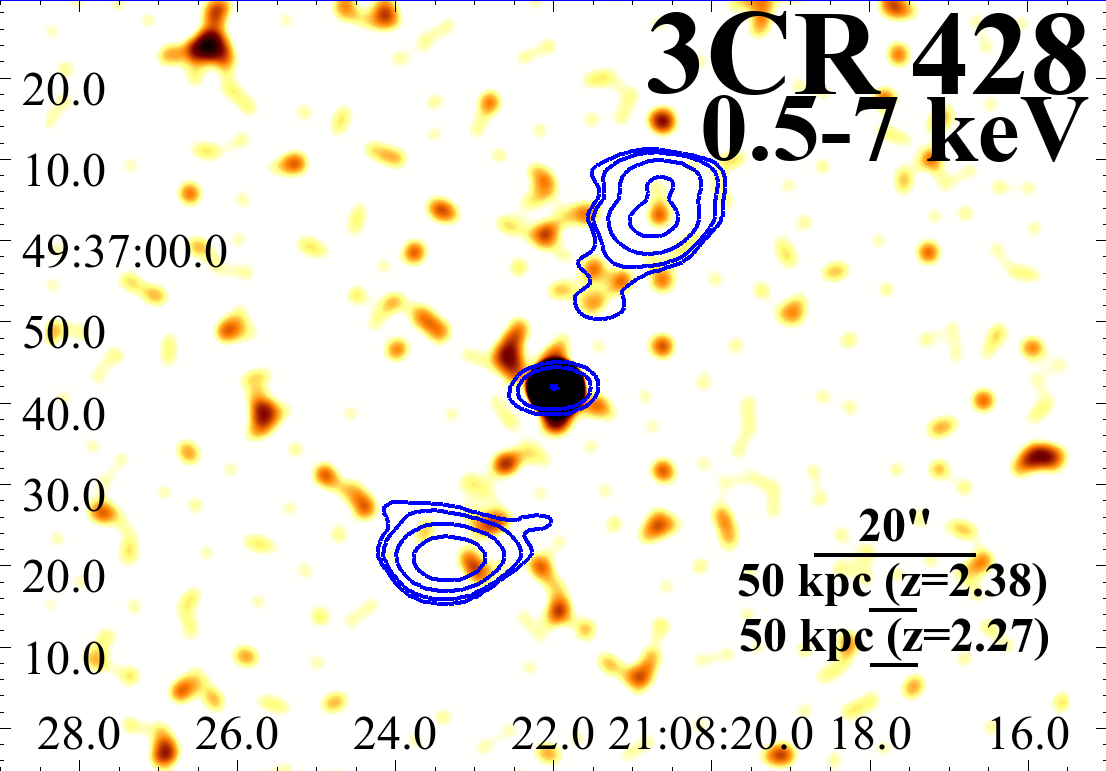}
\caption{(Upper panel): WISE 3.4~$\mu$m filter image of 3CR\,428. VLA contours (black) at 4.5\,GHz are the same used in the \chn\ image (red). Radio contours start from 0.002 Jy/beam and increase by a factor of 2. The arrow points to the IR counterpart of the radio nucleus. No optical counterpart has been detected for this source. (Lower panel): \chn\ X-ray image of 3CR\,428, filtered in the 0.5-7 keV band. Image has not been rebinned, but smoothed with a 6 pixel Gaussian kernel (equivalent to 2.952\arcsec). { In the bottom right of the image, kpcscale measured using the photometric redshifts obtained using the method described in \citet{2017arXiv170908634G} are indicated. }}
\label{fig:3c428chn}
\end{figure}

\begin{figure}[ht!]
\centering
\includegraphics[scale=0.2]{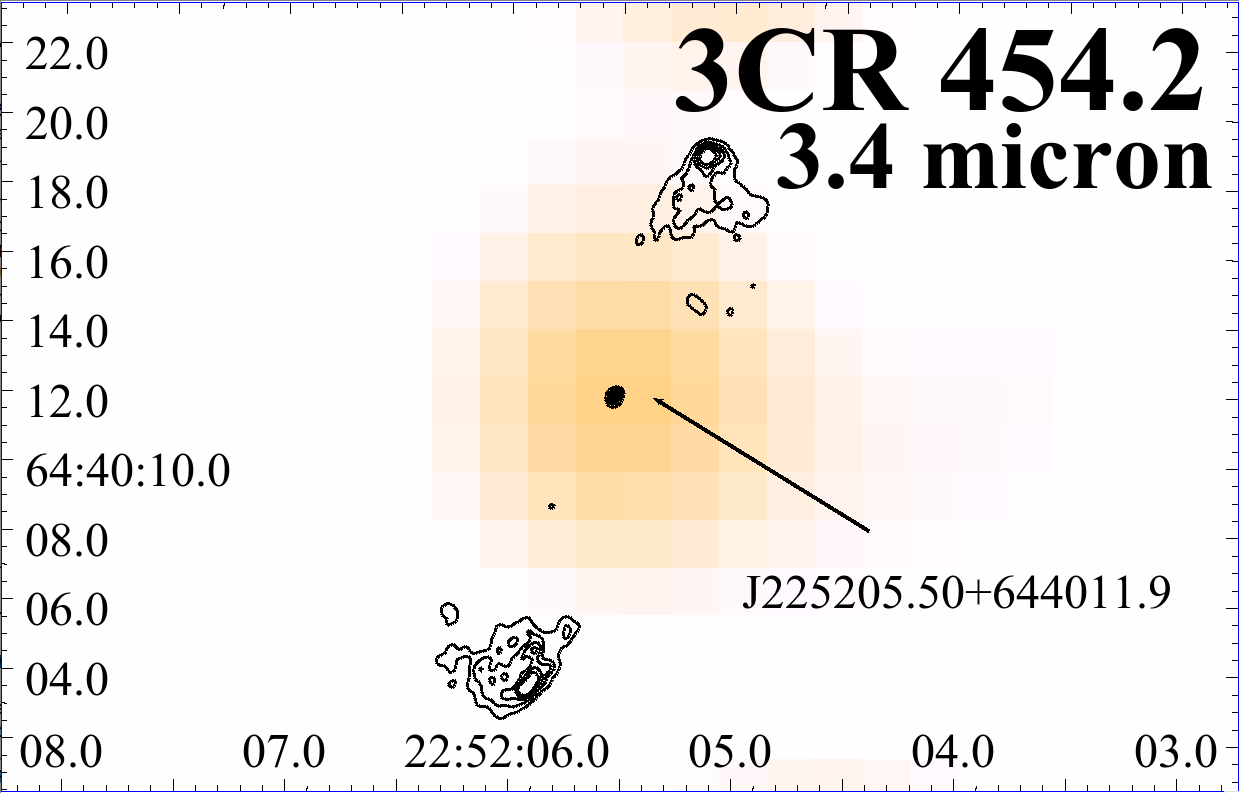}
\includegraphics[scale=0.2]{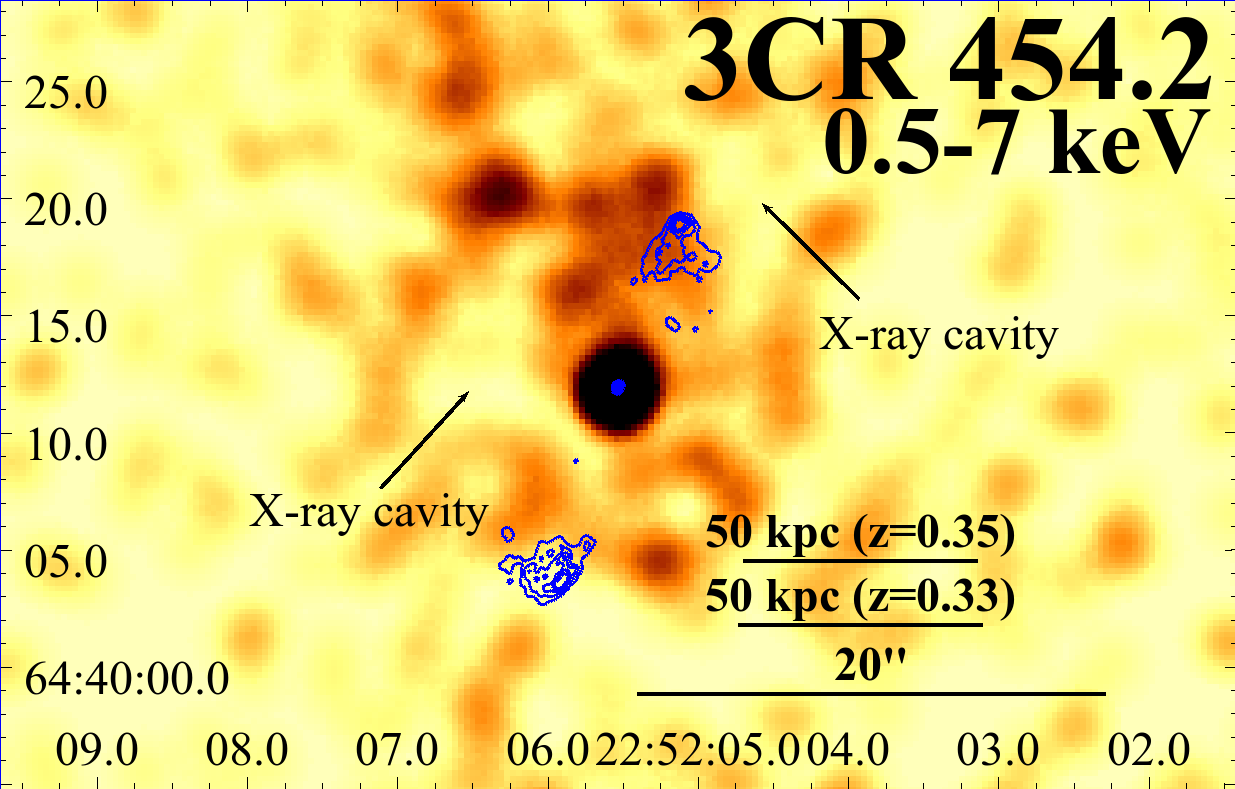}
\caption{(Upper panel): WISE 3.4~$\mu$m filter image of 3CR\,454.2. VLA contours (black) are the same used in the \chn\ image (blue) and start from 0.001 Jy/beam and increasing by a factor of 0.001 Jy/beam to 0.005 Jy/beam. The arrow points to the IR counterpart of the radio nucleus. No optical counterpart has been detected for this source. (Lower panel): \chn\ X-ray image of 3CR\,454.2, filtered in the 0.5-7 keV band, binned up to 0.246 arcsec/pixel and smoothed with a 8 pixels Gaussian kernel (equivalent to 1.968 arcsec). Cavities are indicated by arrows. { In the bottom right of the image, kpcscale measured using the photometric redshifts obtained using the method described in \citet{2017arXiv170908634G} are indicated. }}
\label{fig:3c454.2chn}
\end{figure}

\begin{figure}[ht!]
\centering
\includegraphics[scale=0.16]{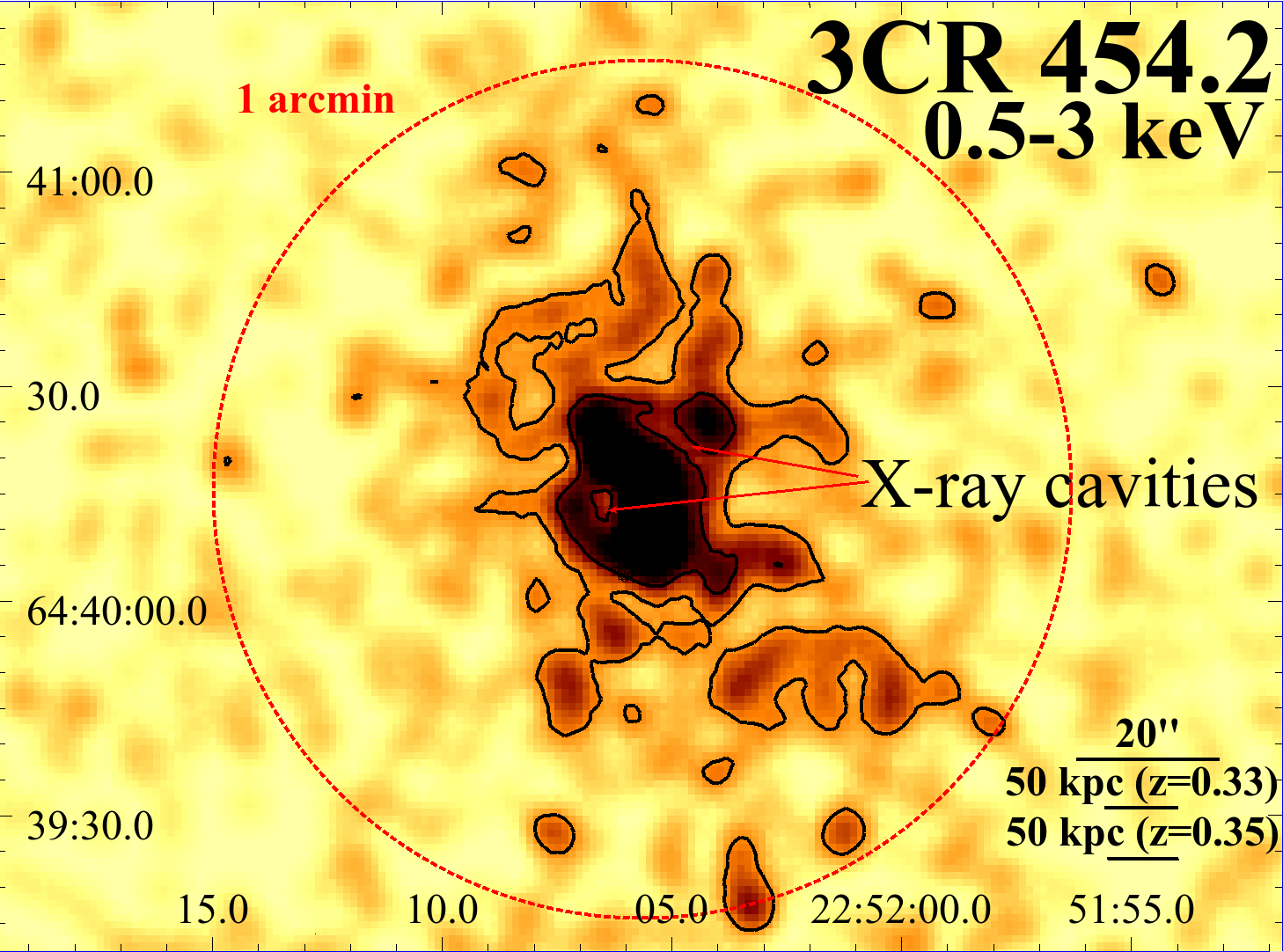}
\caption{\chn\ X-ray image of 3CR\,454.2, filtered in the 0.5-3 keV band, binned up to 0.984 arcsec/pixel and smoothed with a 4.92\arcsec\ Gaussian kernel. The black circle has a 0.5\arcmin radius. The black contours trace the X-ray emission at 0.1 and 0.2 counts/pixel. { In the bottom right of the image, kpcscale measured using the photometric redshifts obtained using the method described in \citet{2017arXiv170908634G} are indicated. }}
\label{fig:3c454.2_ext}
\end{figure}

\begin{figure*}
\centering
\includegraphics[scale=0.25]{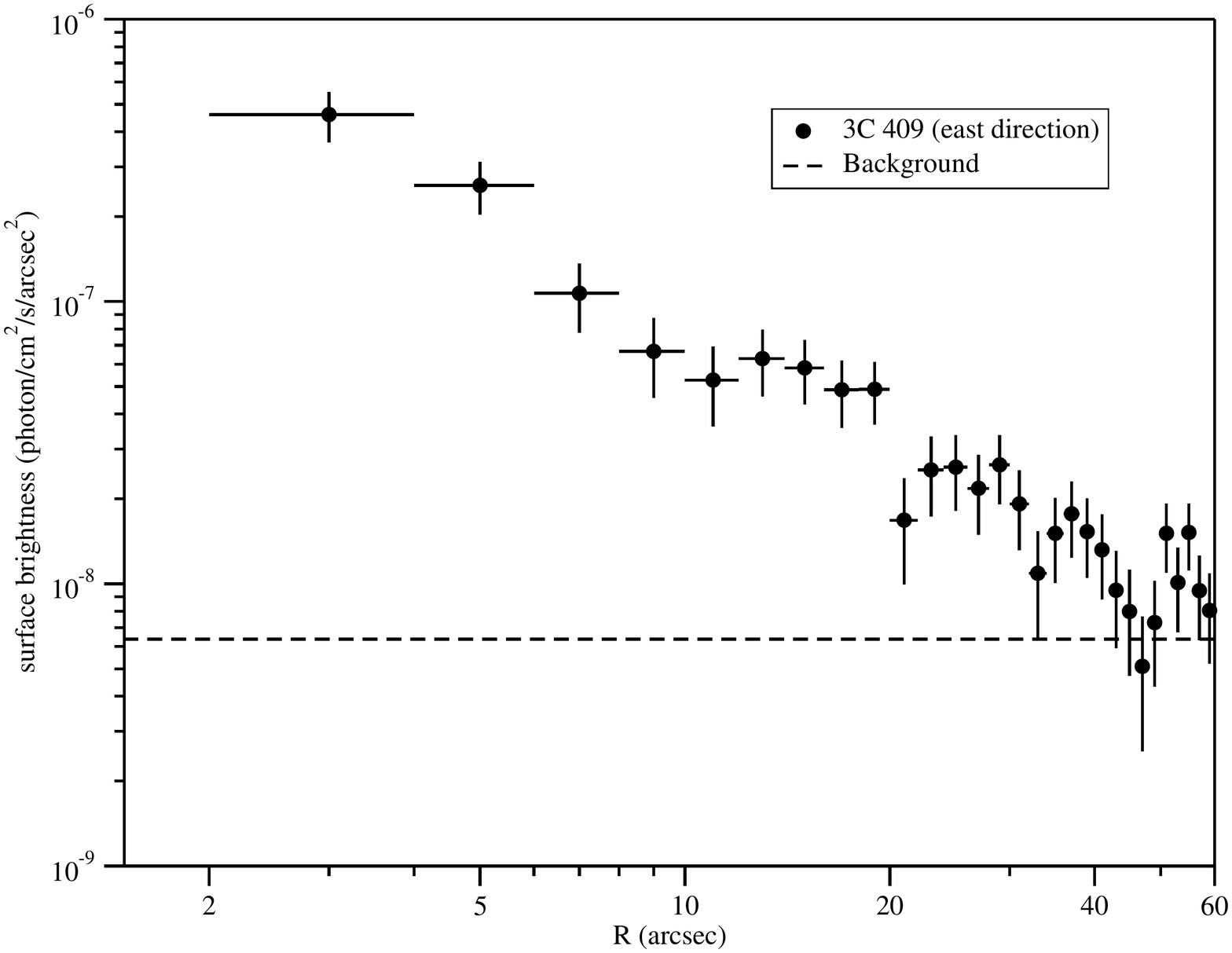}
\includegraphics[scale=0.25]{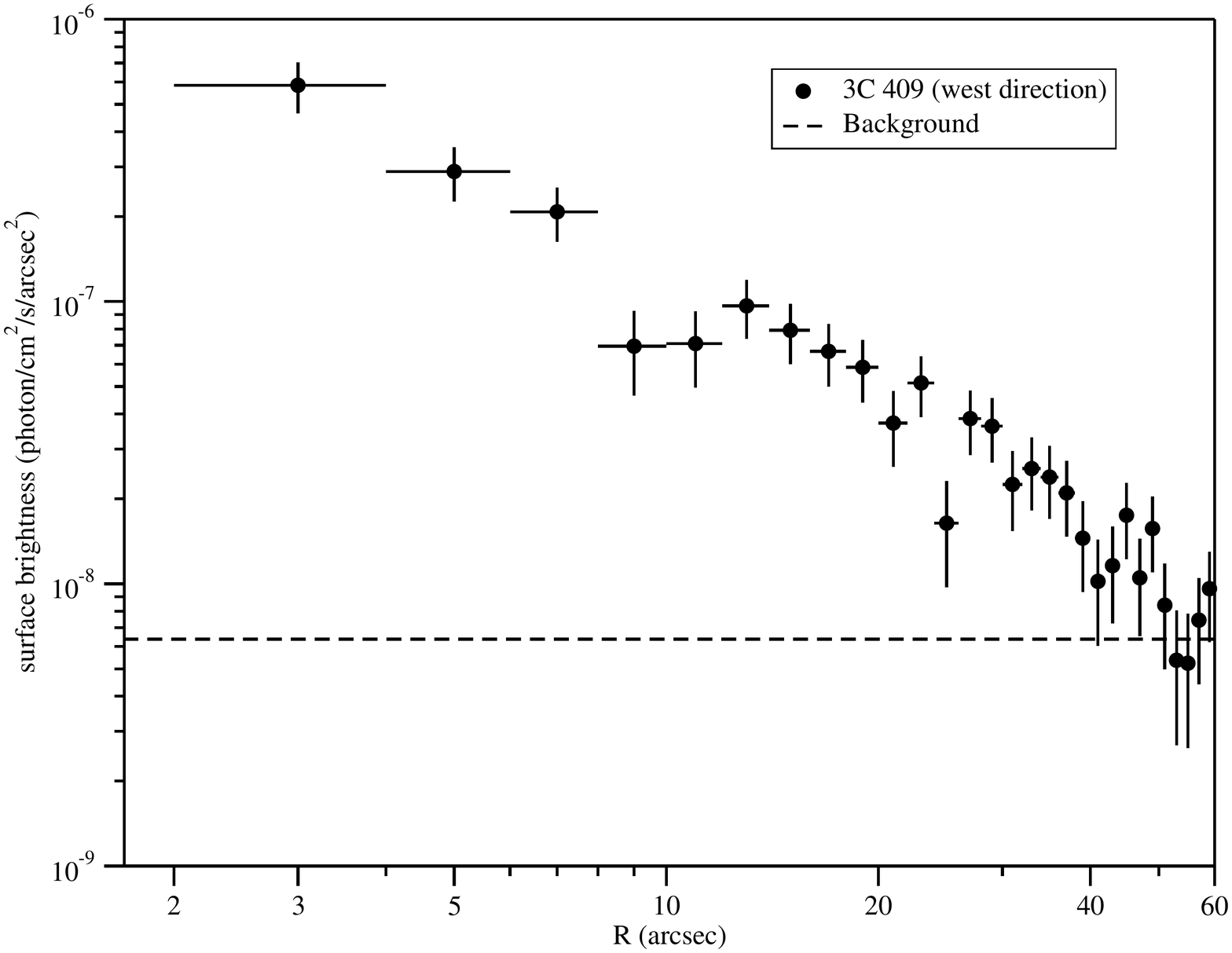}
\includegraphics[scale=0.25]{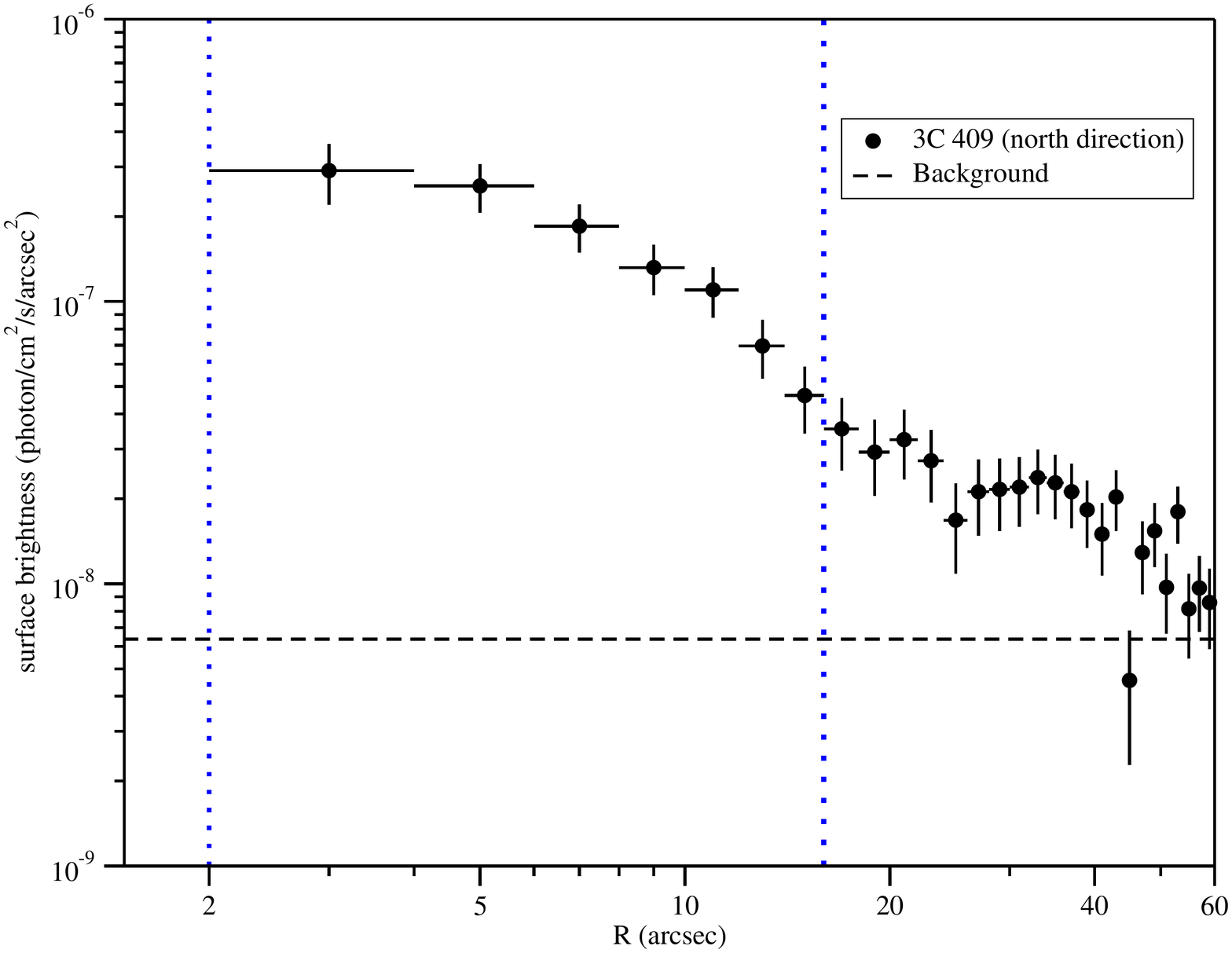}
\includegraphics[scale=0.25]{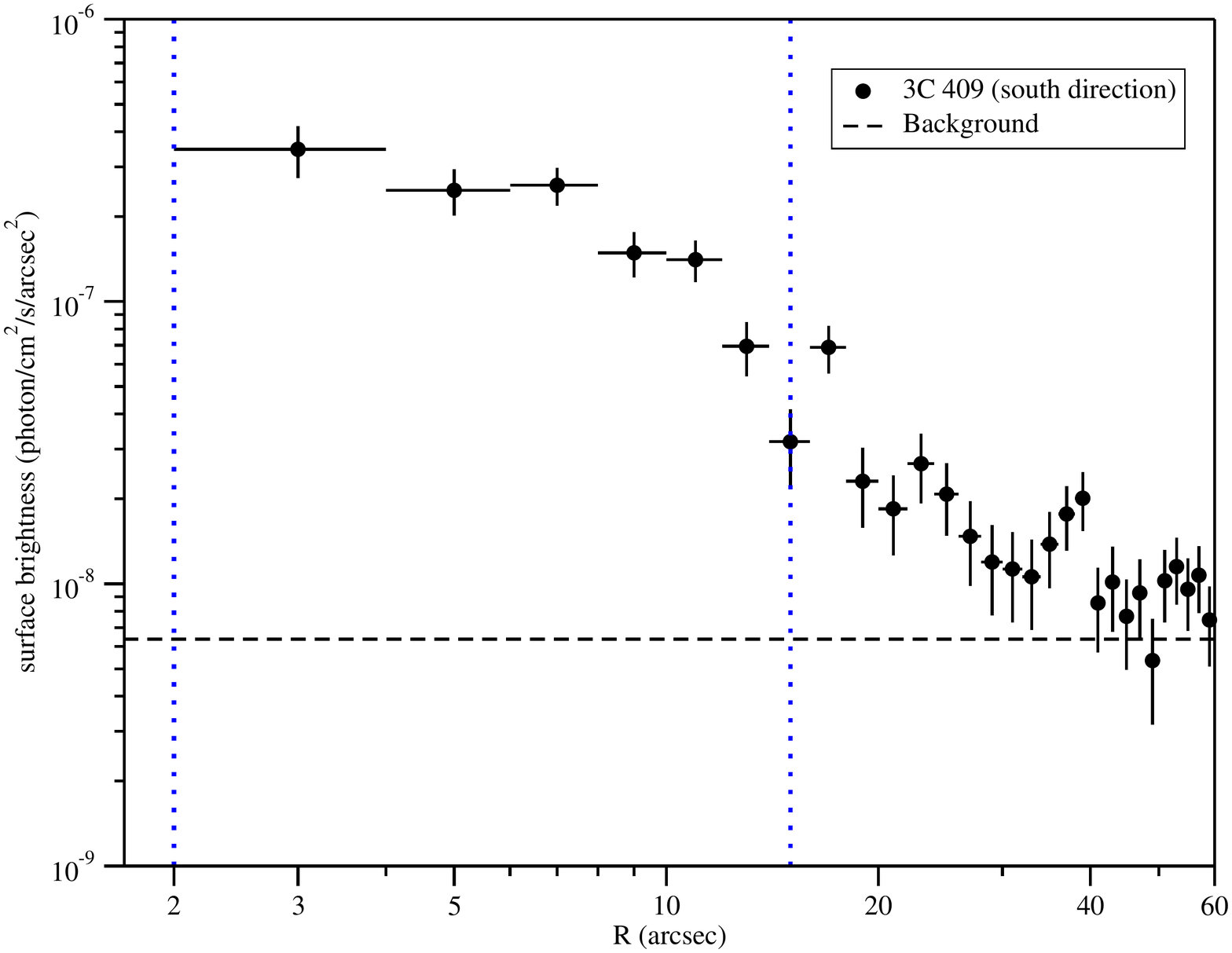}
\caption{Surface brightness profiles for 3CR\,409 extracted in the directions shown in Fig.~\ref{409_sur_bri_regions}. Sectors are divided in bins of $2\arcsec$ width. The inner and outer radii of the lobes are indicated with blue vertical dotted lines. In the western profile (upper right panel) we estimated a jump in the surface brightness between the third and fourth annulus, with 2.8$\sigma$.}
\label{sur_bri}
\end{figure*}

\begin{figure*}
\centering
\includegraphics[scale=0.25]{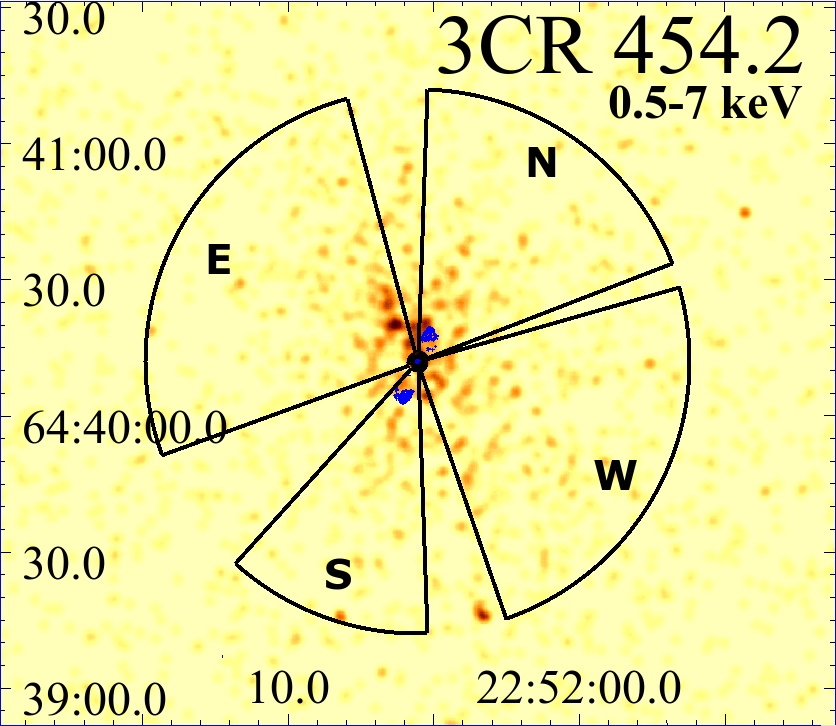}
\caption{Directions used in source 3CR\,454.2 to extract the surface brightness profiles shown in Fig.~\ref{sur_bri454.2}. The four sectors extend up to $60\arcsec$ from the core (that we have excluded, starting from a distance of $2\arcsec$ from the position of the core, as reported in the NVSS, see Table \ref{tab:log}). In blue we show the VLA contours from Fig.~\ref{fig:3c454.2chn}.}
\label{sur_bri_regions}
\end{figure*}

\begin{figure*}
\centering
\includegraphics[scale=0.25]{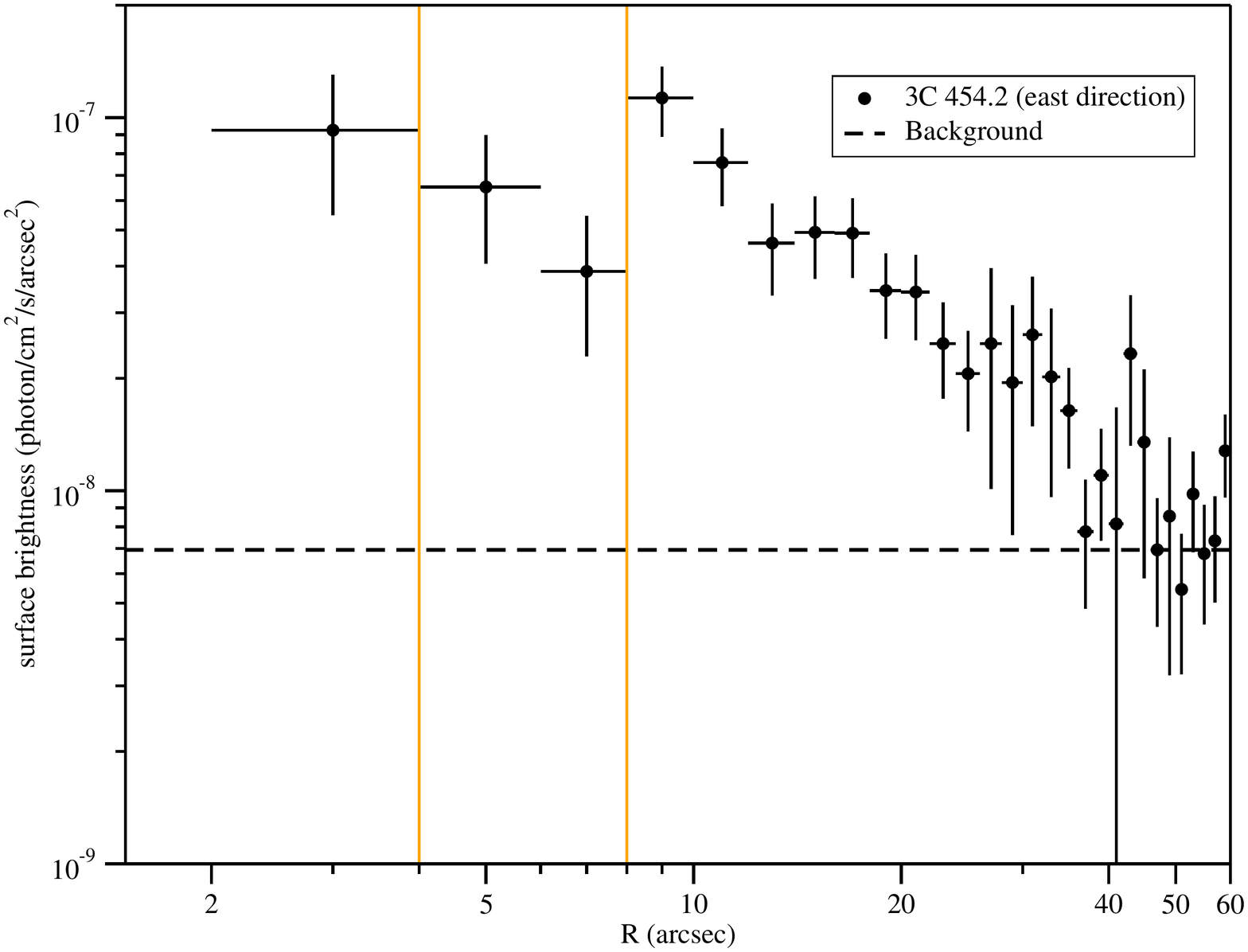}
\includegraphics[scale=0.25]{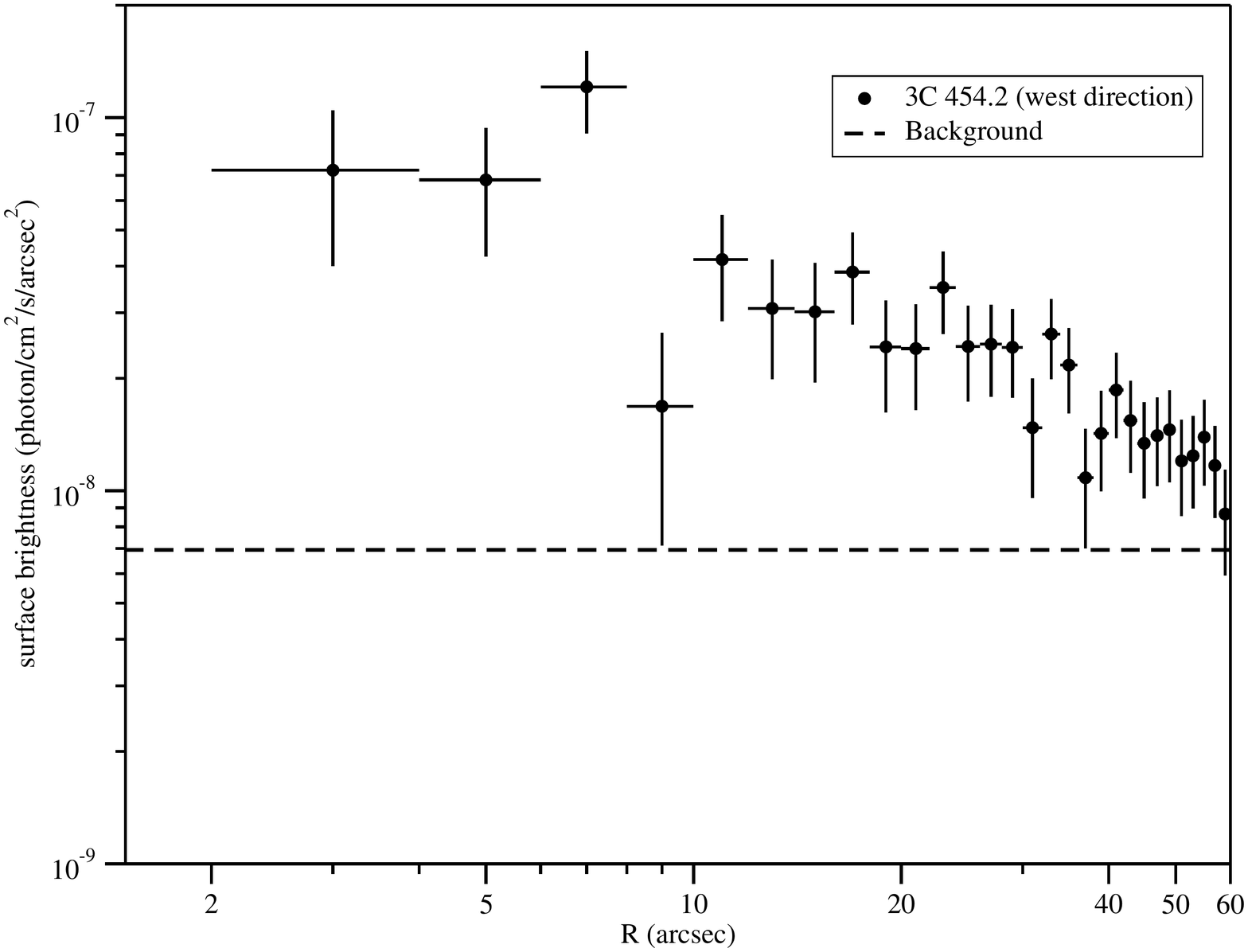}
\includegraphics[scale=0.25]{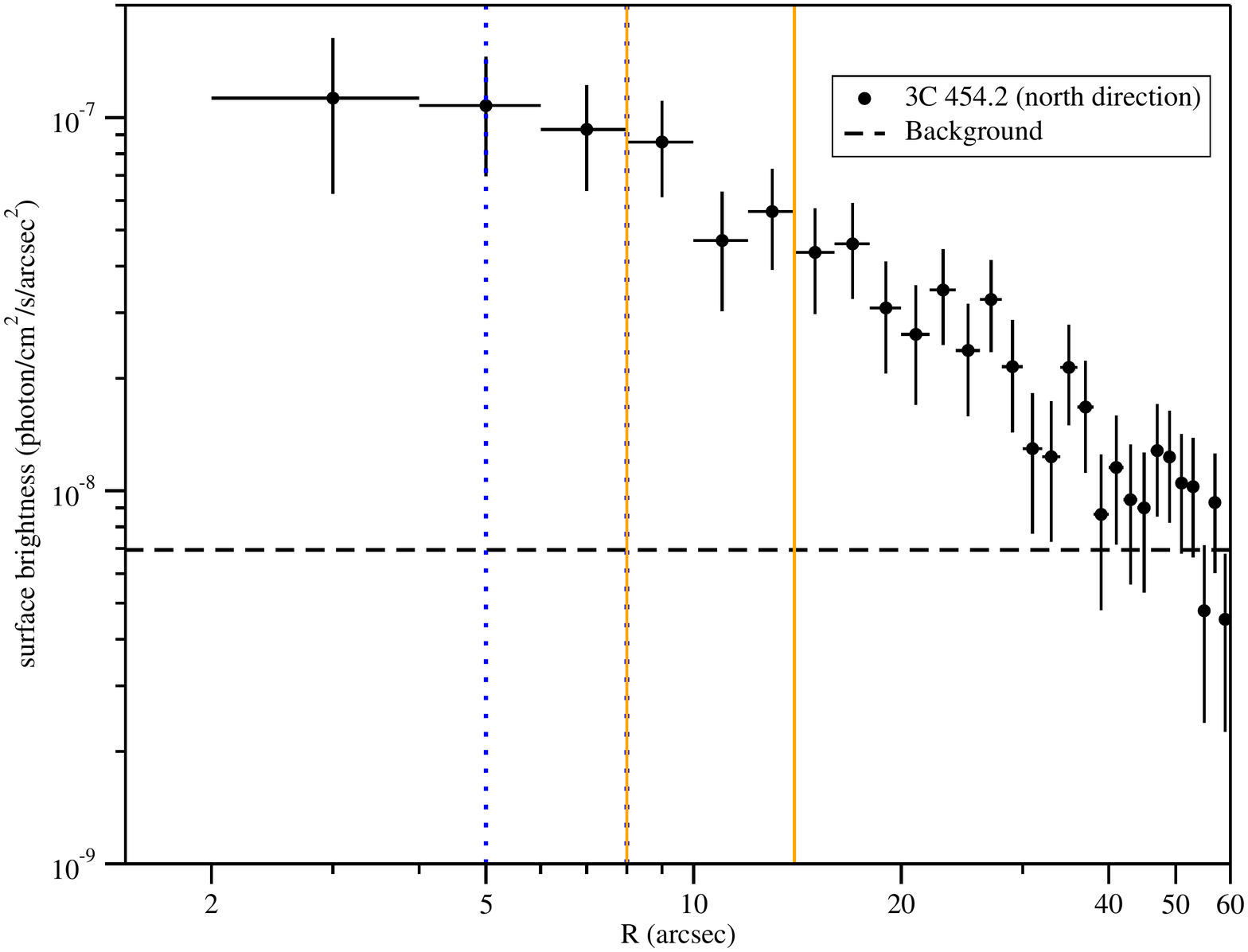}
\includegraphics[scale=0.25]{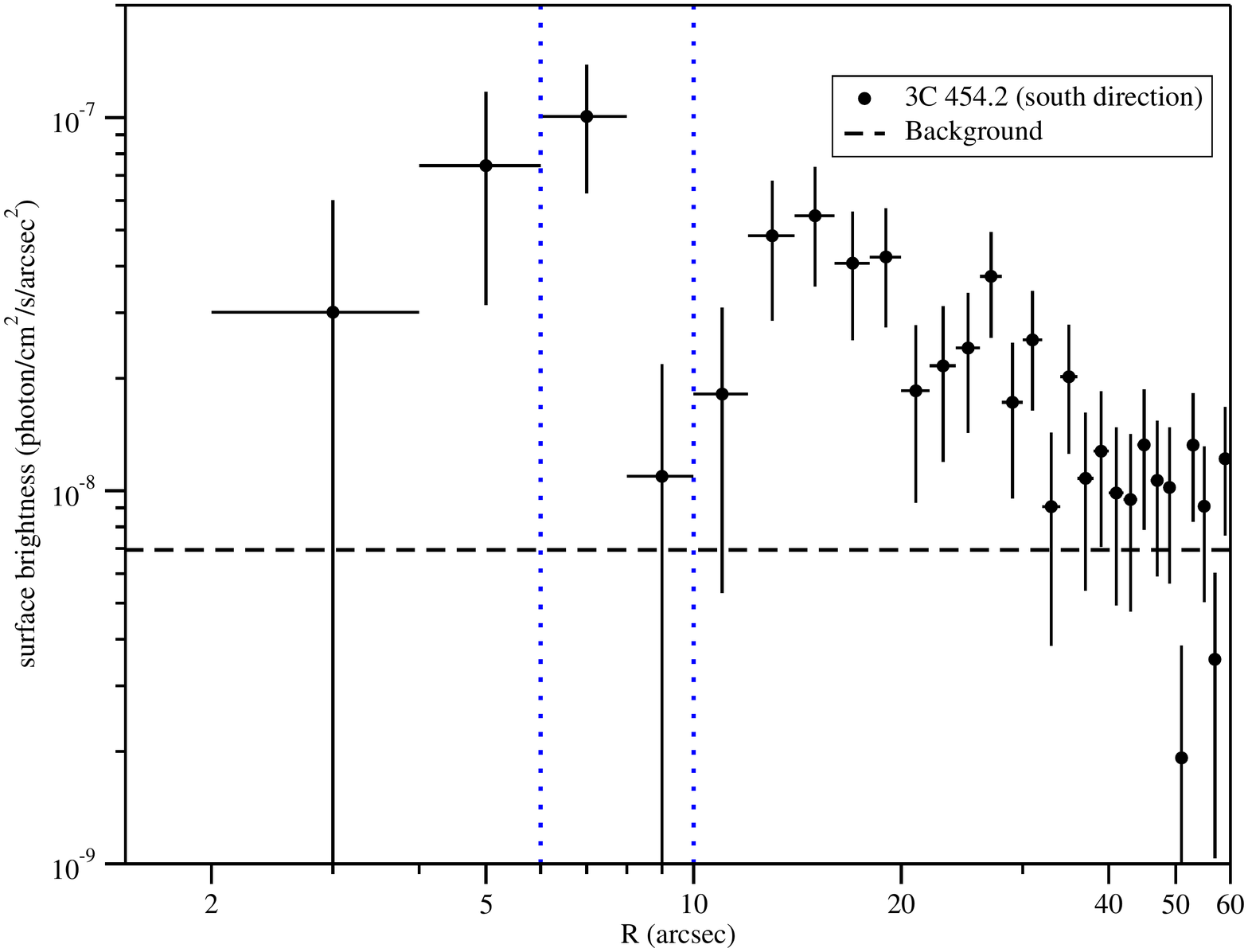}
\caption{Surface brightness profiles for 3CR\,454.2 extracted in the directions shown in Fig.~\ref{sur_bri_regions}. Sectors are divided in bins of $2\arcsec$ width. In the northern and southern profiles (bottom panels of Fig.~\ref{sur_bri454.2}) inner and outer radii of the lobes are indicated with blue vertical dotted lines. In the northern and eastern profiles the areas occupied by the cavities are included between orange vertical lines. }
\label{sur_bri454.2}
\end{figure*}

\subsection{3CR\,428} 
3CR\,428 is a lobe dominated radio source at 4.5\,GHz with a clearly detected core (see black/red contours in Fig.~\ref{fig:3c428chn}, both panels). There is no optical counterpart in the Pan-STARRS R band image of the nucleus while there is a detection at IR frequencies associated with WISE J210822.08+493641.6 located within the XRT positional uncertainty region at an angular separation of 0.5\arcsec. The core is also clearly detected in the \chn\ image, but there are no clear counterparts of radio lobes or hotspots. The X-ray source XRT J210822.1+493642 also matches the NVSS J210822+493637 position at an angular separation of 5.6\arcsec.

As for 3CR\,91, 3CR\, 390, and 3CR\, 409, the WISE counterpart has been recently included in the all-sky catalogue of blazar candidates of \citet{2014ApJS..215...14D}. { Adopting the procedure described in \citet{2017arXiv170908634G}, using the 3.4~$\mu$m WISE magnitude we obtained a photometric redshift of $z=2.38^{+1.62}_{-1.66}$, and using the 4.6$\mu$m WISE magnitude a photometric redshift of $z=2.27^{+1.59}_{-1.58}$, with a probability of 70\% for the source being a QSO.}

Since we detected more than 400 photons in the nuclear region of the \chn\ image, we also performed a spectral analysis of the X-ray core, { adopting an absorbed power-law model. As in the case of 3CR\,409 we obtained a value of $N_{H,int}$ \(\sim {10}^{23} \text{ cm}^{-2}\), explaining our non-detection of an optical counterpart.}


\subsection{3CR\,454.2} 
\label{454.2}
3CR\,454.2 is a lobe dominated radio source in the 8\,GHz VLA image, in which we clearly detected the core and two lobes and hotspots (see black contours in the upper panel of Fig.~\ref{fig:3c454.2chn}). In the Pan-STARRS image there is no optical counterpart located at the radio core position.

In \citet{2016MNRAS.460.3829M} a soft X-ray source XRT J225205.2+644013 was detected, at  an angular separation of 4.6\arcsec\ from the coordinates of NVSS J225205+644010 within its 3CR positional uncertainty region. At an angular separation of 2.3\arcsec\ from this NVSS source, the IR source WISE J225205.50+644011.9 was also found in the AllWISE Catalogue, being its potential counterpart. It is well detected in all filters but the 22~$\mu$m filter. {  Adopting the procedure described in \citet{2017arXiv170908634G}, using the 3.4~$\mu$m WISE magnitude, we obtained a photometric redshift value of $z=0.35^{+0.12}_{-0.11}$, and using the 4.6~$\mu$m WISE magnitude a value of $z=0.33^{+0.12}_{-0.10}$, with a probability of \(96\%\) for the source of being a LERG. }

In the \chn\ image, we highlight the presence of extended X-ray emission and the possible presence of at least two cavities (reported in the lower panel of Fig.~\ref{fig:3c454.2chn} and, at a larger scale, in Fig.~\ref{fig:3c454.2_ext}). As in the case of 3CR\,409 we have derived the surface brightness profiles in the directions shown in Fig.~\ref{sur_bri_regions}. Again, northern and southern directions have been selected to encompass the radio lobes contours, while eastern and western directions are that away from the lobes. We found evidence for diffuse emission up to $\sim$ 50\arcsec\ from the core (see Fig.~\ref{sur_bri454.2}). The northwestern cavity has less than 2$\sigma$ level significance, while the southeastern one has a significance of 3.4$\sigma$. To evaluate the significance of the cavities, we have estimated the counts in each cavity, using circular regions of appropriate radii, and the average level of the diffuse emission at the same distance from the core using several circular regions with the same radii of the cavity regions. Then, using Poisson statistics, we evaluated the Gaussian $\sigma$ equivalent of the cavities significance. We have performed a spectral analysis of the diffuse X-ray emission, adopting a thermal model, in every possible combination of redshift and abundance values, fixed or free.{  As specified in Subsection~\ref{subsec:spec}, we have excluded a 2\arcsec circular region including the nuclear emission, but since the contribution of the wings of the PSF was lower than 1\%, we have excluded this contribution in the thermal model of the extended emission. Due to low count statistics, all fits are poorly constrained and the uncertainties on the temperature are large. Also in this case, deeper \chn\ observations are needed to properly constrain the properties of the cluster IGM.}

\section{Summary and Conclusions}
\label{sec:summary}
In this paper, we present a multi-wavelength (radio, infrared, optical and X-ray) study of seven of the 25 extragalactic radio sources listed in the Third Cambridge Revised Catalog (3CR) as unidentified by \citep{1985PASP...97..932S}. All these sources, { previously,} lacked a confirmed optical counterpart and thus miss redshift and optical classification.
The 3CR \chn\ Snapshot Survey, started in 2008, aimed at searching for X-ray emission from jet knots, hotspots and lobes, studying the nuclear emission of their host galaxies and investigating their environments at all scales. 

Adopting the same procedures used in the previous papers of the Snapshot, we can summarize our results as follows:
    \begin{itemize}
    \item six of the seven sources (all but 3CR\,409), show a clear detection of the radio core at 1.5\,GHz, 6\,GHz and 10 GHz. These radio images, retrieved from the historical VLA archive, were all manually reduced { and for five sources we also gave a tentative FRII radio classification;}
    \item we found IR counterparts to all the radio cores, thanks to the WISE archival images. This allowed us to estimate the photometric redshift of the counterparts using the magnitudes at 3.4 and 4.6 $\mu$m { as described in \citet{2017arXiv170908634G}. This method allowed us also to give tentative classifications (LERG/HERG/QSO) of the sources in the sample. Most of the sources are classified as QSOs with a probability $\gtrsim$ 60\% , while 3CR\,131 and 3CR\,454.2 are classified as LERGs, with the same probability;}
    \item only three sources (namely 3CR\,91, 3CR\,158 and 3CR\,390) of the seven with an infrared counterpart are also detected
    in the optical band using Pan-STARRS images. { For the other sources we have obtained an $N_{H,int}$ value of the order of \(\sim {10}^{23} \text{ cm}^{-2}\), and corresponding levels of dust obscuration are likely the reason for the non-detection of the optical counterpart};
    \item we found \chn\ X-ray counterparts for all the radio cores. Then, for 3CR\,91, 3CR\,390, 3CR\,490 and 3CR\,428, we also estimated the { X-ray spectral indeces (\(\alpha_X\)=0.48-0.80)} and the intrinsic absorption $N_{H,int}$, via spectral analysis. { The spectral indeces are compatible with results reported in the literature for the nuclei of QSOs and LERGs};
    \item we detected X-ray emission arising from the X-ray counterpart of the northern radio jet in 3CR\,158 as well as that associated with the radio bridge in 3CR\,390;
    \item our \chn\ observations revealed also the presence of extended X-ray emission, the hallmark of galaxy clusters, around 3CR\,409 and 3CR\,454.2. We performed a spectral analysis, but temperature or spectral parameters are unconstrained. { This demands deeper \chn\ observations to make more conclusive measures of the temperature, mass, and luminosity of the clusters}.
\end{itemize}

{ Regarding the tentative classification of the sources in our sample, our results are:
\begin{itemize}
	\item 3CR\,91 has an 8 GHz radio structure similar to an FRII radio galaxy. From the WISE magnitudes of the counterpart it can be classified as a QSO at $z\sim0.2$. The Pan-STARRS core counterpart is detected, and the intrinsic absorption value estimated from X-ray spectral fitting is of the same order of magnitude as the Galactic one. We therefore classify this source as a QSO at $z\sim0.2$.
	\item 3CR\,131 has an 8 GHz radio structure similar to an FRII radio galaxy. From the WISE magnitudes of the counterpart, it can be classified as a LERG at $z\sim0.4$. There is no Pan-STARRS detected counterpart, but there are not enough nuclear counts in the \chn\ image to perform a spectral analysis. We classify this source as an FRII-LERG at $z\sim0.4$.
	\item 3CR\,158, at 8 GHz, is similar to an FRII, and we detected X-ray extended emission aligned with the radio jet structure. The photometric redshift estimate from WISE magnitudes is $z\sim 4$, and the source is classified as a QSO. The poor statistics do not allow us to perform a spectral analysis of the source nucleus. The Pan-STARRS core counterpart is detected with \(m_R = 20.6\), the highest of the sample. From these results, we classify this source as a high redshift (\(\gtrsim 4\)) QSO.
	\item 3CR\,390 does not have a clear radio morphology, showing two sided lobes. From the WISE magnitudes of the counterpart, this source can be classified as a QSO at $z\sim0.4$. This source has a detected Pan-STARRS nuclear counterpart, and from the X-ray spectral fitting, we estimate a nuclear intrinsic absorption comparable to the Galactic one. We classify this source as a QSO at \(z\sim 0.4\).
	\item 3CR\,409 shows an FRII radio morphology, and from the WISE magnitudes this source is classified as a QSO at $z\sim0.9$. This source appears to lie in a cluster of galaxies, and does not have a detected Pan-STARRS nuclear counterpart. From the X-ray spectral fitting, we estimate an intrinsic absorption \(\sim {10}^{23} \text{ cm}^{-2}\). We therefore classify this source as a highly-absorbed QSO at \(z\sim 0.9\).
	\item 3CR\,428 features an FRII-like radio morphology. The WISE magnitudes of this source classify it as a QSO at $z\sim 2.3$. No Pan-STARRS nuclear counterpart is detected, and from the X-ray spectral fit we estimate an intrinsic absorption \(\sim {10}^{23} \text{ cm}^{-2}\). We classify this source as a highly-absorbed QSO at \(z\sim 2.3\).
	\item 3CR\,454.2 features an FRII-like radio morphology. From the WISE magnitudes we classify this source as a LERG at $z\sim 0.3$. We do not detect a nuclear Pan-STARRS counterpart. This source appears to lie in a galaxy cluster with disturbed IGM morphology, however the available \textit{Chandra} data do not allow nuclear spectral analysis and so no estimate of the intrinsic absorption, or the IGM temperature. Deeper X-ray observations are needed to draw firm conclusions on this source. 
	\end{itemize}

In conclusion, we have four sources with a predicted low-z host ($z\sim 0.2-0.4$) namely 3CR\,91, 3CR\,131, 3CR\,390, 3CR\, 454.2. Two sources are classified as high-z ($ > 0.9$), highly-absorbed QSO, namely 3CR\,158, 3CR\,409 and 3CR\,428. In the case of 3CR\,158, since the estimated photometric redshift is that of a very high-z source, deeper \chn\ observations are required to verify this estimate. 

}

This paper presents the first attempt to describe in more detail the unidentified sources in the 3CR Catalog, using new \chn\ X-ray observations and archival observations from VLA, WISE and Pan-STARRS observatories, leading to the discovery of X-ray emission from nuclei, jets, and cluster gas. This last part of the 3CR Snapshot Survey, devoted to the 3CR unidentified sources, is still ongoing and we expect to close this exploratory sample of nine sources - awarded in \text{Chandra} observation Cycle 21 - around April 2021, with additional sources scheduled for the rest of the year.

\acknowledgments 
{ We thank the referee for a careful reading of our manuscript and many helpful comments that led to improvements in the paper.}
This work is supported by the ``Departments of Excellence 2018 - 2022'' Grant awarded by the Italian Ministry of Education, University and Research (MIUR) (L. 232/2016). This research has made use of resources provided by the Ministry of Education, Universities and Research for the grant MASF\_FFABR\_17\_01. This investigation is supported by the National Aeronautics and Space Administration (NASA) grants GO9-20083X and GO0-21110X. A.P. acknowledges financial support from the Consorzio Interuniversitario per la fisica Spaziale (CIFS) under the agreement related to the grant MASF\_CONTR\_FIN\_18\_02. A.J. acknowledges the financial support (MASF\_CONTR\_FIN\_18\_01) from the Italian National Institute of Astrophysics under the agreement with the Instituto de Astrofisica de Canarias for the ``Becas Internacionales para Licenciados y/o Graduados Convocatoria de 2017''. W.F.  and R.K. acknowledge support from the Smithsonian Institution and the Chandra High Resolution Camera Project through NASA contract NAS8-03060. C.S. acknowledges support from the ERC-StG DRANOEL, n. 714245.
The National Radio Astronomy Observatory is operated by Associated Universities, Inc., under contract with the National Science Foundation.
This research has made use of data obtained from the High-Energy Astrophysics Science Archive Research Center (HEASARC) provided by NASA's Goddard Space Flight Center; 
the SIMBAD database operated at CDS, Strasbourg, France; the NASA/IPAC Extragalactic Database (NED) operated by the Jet Propulsion Laboratory, California
Institute of Technology, under contract with the National Aeronautics and Space Administration.
TOPCAT\footnote{\underline{http://www.star.bris.ac.uk/$\sim$mbt/topcat/}} \citep{2005ASPC..347...29T} for the preparation and manipulation of the tabular data and the images.
SAOImage DS9 were used extensively in this work for the preparation and manipulation of the images.  SAOImage DS9 was developed by the Smithsonian Astrophysical Observatory. 

{Facilities:} \facility{VLA}, \facility{CXO (ACIS), \facility{WISE}, \facility{Pan-STARRS}.}

\begin{table*} 
\caption{Summary of X-ray observations}
\label{tab:log}
\begin{center}
\begin{tabular}{|lrrrrrrrrrr|}
\hline
3CR name &  R.A. (J2000) & Dec. (J2000) & N$_{H,Gal}$ & \chn & Obs. date & Exposure & S$_{178}$ & Counterpart & Radio nucleus & Cluster   \\
 & (hh mm ss)   & (dd mm ss)    & (10$^{21}$cm$^{-2}$) & Obs. ID  & yyyy-mm-dd & (ks) & (Jy) & & &  \\ 
\hline 
\noalign{\smallskip}
91 & 03 37 43.032 & +50 45 47.622 & 4.88 & 22626 & 2019-11-18 & 18.9 & 14.1 & IR, opt & yes & no  \\
131 & 04 53 23.337 & +31 29 27.826 & 2.36 & 22627 & 2019-12-29 & 23.8 & 14.6 & IR & yes & no  \\
158 & 06 21 41.041 & +14 32 13.035 & 4.95 & 22629 & 2020-01-10 & 23.75 & 18.1 & IR, opt & yes & no   \\
390 & 18 45 37.601 & +09 53 44.998 & 3.00 & 22630 & 2020-02-28 & 19.8 & 21.0 & IR, opt & yes & no  \\
409 & 20 14 27.74* & +23 34 58.4* & 2.49 & 22631 & 2019-11-29 & 19.81 & 76.6 & IR & no & yes  \\
428 & 21 08 21.985 & +49 36 41.820 & 10.9 & 22632 &  2019-12-23 & 20.79 & 16.6 & IR & yes & no \\
454.2 & 22 52 05.530 & +64 40 11.940 & 7.48 & 22633 & 2019-11-17 & 19.8 & 8.8 & IR & yes & yes\\

\noalign{\smallskip}
\hline
\end{tabular}\\
\end{center}
Col. (1): The 3CR name.
Col. (2-3): The celestial positions obtained from the radio images (the only exception is 3CR\, 409 where we used the NVSS counterpart coordinates).
Col. (4): Galactic Neutral hydrogen column densities N$_{H,Gal}$ along the line of sight \citep{2005A&A...440..775K}.
Col. (5): The \chn\ observation ID.
Col. (6): The date when the \chn\ observation was performed.
Col. (7): Exposure time in ks, as reported in the \chn\ Archive.
Col. (8): S$_{178}$ is the flux density at 178 MHz, from \citet{1985PASP...97..932S}.
Col. (9-11): Remarks of the results of this work.

\end{table*}

\clearpage
\begin{table*} 
\caption{Summary of radio observations}
\label{tab:radio_obs}
\tiny
\begin{center}
\begin{tabular}{|lllllllllll|}
\hline
3CR Name & NRAO ID & Obs. Date & Frequency & Configuration & Beam size & Total Flux & Peak Flux & TOS & RMS ($\sigma$) & contour levels \\
&    &  yyyy-mm-dd & (GHz) & & (${arcsec}^2$) & (Jy) & (Jy/beam) & (s)  & (10$^{-3}$Jy/beam) & ($\sigma$)\\ 
\hline 
\noalign{\smallskip}
  91 & AH976 & 2008-10-09 & 8.44, 8.49 & AB & 1 $\times $ 0.5 & 0.63$\pm$0.06 & 4.25$\times10^{-2}$ & 70 & 1.4 & 6, 8, 16, 32, 64, 72, 74, 78\\
  - & AP001  & 1986-09-15 & 1.45 & B & 5.1 $\times $  4.2 & 3.39$\pm$0.17 & 1.43 & 150 & 1.23 & 9, 27, 81, 243, 729, 1162\\
  131 & AH976 & 2008-10-09 & 8.44, 8.49 & A & 0.28 $\times $ 0.22 & 0.36$\pm$0.04 & 1.28 & 70 & 0.33 &  5, 20, 80, 320, 1280, 5120\\
  - & AP001 & 1986-09-15 & 1.45 & B & 7 $\times$ 6 & 2.8$\pm$0.14 & 1.3 & 150 & 2.8 & 4, 8, 16, 32, 64, 128, 256, 512\\
  158 & AH976 & 2008-10-09 & 8.44, 8.49 & A & 0.22 $\times $ 0.20 & 0.63$\pm$0.03 & 1.28$\times10^{-1}$ & 70 & 0.44 & 3, 9, 27, 81, 243\\
  - & AR069 & 1983-10-13 & 4.84, 4.89 & A & 0.5 $\times$  0.4 & 0.58$\pm$0.03 & 1.57$\times10^{-1}$ & 260 & 0.29 & 3, 6, 12, 24, 48, 96, 192, 384, 548 \\
  - & AF156 & 1989-07-21 & 4.84, 4.89 & BC & 5.52 $\times$ 4.17 & 0.58$\pm$0.03 & 3.93$\times10^{-1}$ & 310 & 0.17 & 5, 9, 27, 81, 243, 729, 2187, 2258\\
  390 & AT147 & 1993-03-24 & 4.84, 4.89 & B & 1.3 $\times$ 1.2 & 1.17$\pm$0.06 & 3.04$\times10^{-1}$ & 370 & 0.10 & 10, 40, 160, 640, 2560\\
  - & AK100 & 1984-02-27 & 1.41, 1.64 & B & 4.1 $\times$ 3.5 & 4.42$\pm$0.22 & 2.145 & 400 & 0.69 & 8, 16, 32, 64, 128, 256, 512, 1024, 2048, 3110 \\
  - & AP001 & 1986-11-28 & 1.45 & C & 23.9$\times$16.6 & 4.6$\pm$0.23 & 2.4$\times10^{-1}$ & 210 & 2.8 & 6, 12, 24, 48, 96, 192, 384, 768, 1536 \\
  409 & AP001 & 1986-09-15 & 1.45 & BC & 3.5 $\times$ 3.1 & 12.54$\pm$0.63 & 2.06$\times10^{-1}$ & 190 & 4 & 5, 20, 80, 320, 1280\\
  - & AC169 & 1986-08-09 & 1.39, 1.42, 1.46, 1.51, 1.63, 1.66 & B & - & - & - & 980 & - & -\\
  428 & AF102 & 1985-07-30 & 4.84, 4.89 & C & 6.4 $\times$ 3.5 & 0.51$\pm$0.02 & 2.06 & 300 & 0.26 & 2, 4, 8, 16, 32, 64, 128, 512, 1024, 8000  \\
  - & AH147 & 1984-09-19 &1.45, 1.5 & D & 43.2 $\times$ 38 & 2.16$\pm$0.11 & 1.17 & 4680 & 0.45 & 6, 12, 24, 48, 96, 192, 384, 768, 1536, 3072\\
  454.2 & AH976 & 2008-10-09 & 8.44, 8.49 & A & 0.34 $\times$ 0.28 & 0.25$\pm$0.02 & 1.11$\times10^{-2}$& 80 & 0.29 & 3, 6, 12, 24, 38 \\
  - & AS238 & 1985-07-14 & 4.84, 4.89 & C & 5.7 x 4 & 0.66$\pm$0.03 & 1.82$\times10^{-1}$ & 270 & 0.27 & 6, 12, 24, 48, 96, 192, 384, 674 \\
\noalign{\smallskip}
\hline
\end{tabular}\\
\end{center}
Col. (1): The 3CR name.
Col. (2): The NRAO observing project (or proposal) identification.
Col. (3): Date of the observation.
Col. (4): Frequency of the VLA observation.
col. (5): Array configuration.
Col. (6): Size of the elliptical clean beam (major axis $\times$ minor axis).
Col. (7): Total flux of the source, as obtained from the self-calibration.
Col. (8): Peak flux of the radio image.
Col. (9): Observation Time On Source (TOS).
Col. (10): Root Mean Square (RMS) noise of the clean image.
Col. (11): Contours levels in units of RMS.

\end{table*}

\begin{table*} 
\caption{Summary of optical and IR observations}
\label{tab:optir}
\tiny
\begin{center}
\begin{tabular}{|lcclccccccl|}
\hline
3CR name & NVSS name & WISE name & E(B-V) & w1 & w2 & w3 & R band & $z_{w1}$ & $z_{w2}$ & VLA/\chn\ \\
 & & & (mag) & (mag) & (mag) &  (mag) & (mag) & & & (arcsec) \\ 
\hline 
\noalign{\smallskip}
91 & J033743+504552 & J033743.02+504547.6 & 1.05$\pm$0.05 & 11.885$\pm$0.022 & 10.80$\pm$0.021 & 7.936$\pm$0.020 & 20.29$\pm$0.05 & $0.23\pm0.18$ & $0.19^{+0.18}_{-0.14}$ & 0.43\\
131 & J045323+312924 & J045323.34+312928.4 & 0.83$\pm$0.02 & 14.98$\pm$0.041 & 14.77$\pm$0.082 & 12.306 & - & $0.41^{+0.13}_{-0.12}$ & $0.44\pm0.13$ & 0.65\\
158 & J062141+143211 & J062141.01+143212.8 & 0.77$\pm$0.02 & 15.133$\pm$0.046 & 13.953$\pm$0.043 & 11.131$\pm$0.189 & 20.62$\pm$0.06 & $4.57^{+0.68}_{-3.14}$ & $3.16^{+2.09}_{-2.19}$ & 0.35\\
390 & J184537+09534 & J184537.60+095345.0 & 0.49$\pm$0.01& 12.546$\pm$0.043 & 11.575$\pm$0.024 & 9.150$\pm$0.029 & 18.80$\pm$ 0.06 & $0.41^{+0.28}_{-0.32}$ & $0.30^{+0.21}_{-0.25}$ & 0.02\\
409 & J201427+233452 & J201427.59+233452.6 & 0.57$\pm$0.03 & 13.547$\pm$0.050 & 12.377$\pm$0.027 & 9.005$\pm$0.027 & - & $1.04^{+0.71}_{-0.74}$ & $0.55^{+0.38}_{-0.46}$ & *\\
428 & J210822+493637 & J210822.08+493641.6 & 2.59$\pm$0.07 & 14.559$\pm$0.064 & 13.097$\pm$0.035 & 10.143$\pm$0.056 & - & $2.38^{+1.62}_{-1.66}$ & $2.27^{+1.59}_{-1.58}$ & 1.09\\
454.2 & J225205+644010 & J225205.50+644011.9 & 1.26$\pm$0.04 & 14.652$\pm$0.030 & 14.341$\pm$0.042 & 13.121$\pm$0.467 & - & $0.35^{+0.12}_{-0.11}$ & $0.33^{+0.12}_{-0.10}$  & 0.72\\
\noalign{\smallskip}
\hline
\end{tabular}\\
\end{center}
Col. (1): The 3CR name.
Col. (2): Associated NVSS source. 
Col. (3): Associated WISE source. 
Col. (4): Extinction, as reported in the NASA/IPAC Infrared Science Archive (IRSA).
Col. (5): Magnitude in the WISE w1 filter (3.4 $\mu$m).
Col. (6): Magnitude in the WISE w2 filter (4.6 $\mu$m).
Col. (7): Magnitude in the WISE w3 filter (12 $\mu$m).
Col. (8): Magnitude in the Pan-STARRS R band.
Col. (9): Median redshift value obtained from 3.4~$\mu$m filter magnitude.
Col. (10): Median redshift value obtained from 4.6~$\mu$m filter magnitude.
Col. (11): Angular separation between the position of the radio core detected in the VLA maps and that of the associated X-ray counterpart in the \chn\ images (see Sect.~\ref{sec:xrays}) .
*Since we were unable to detect the position of the radio core of 3CR\,409, the swift is missing.
{** For the  estimate of the photometric redshift we have used dereddened values of the WISE magnitude corrected for Galactic absorption using reddening estimates from \citet{2011ApJ...737..103S} and the extinction model from \citet{2007ApJ...663..320F}.}

\end{table*}

\begin{table*} 
\caption{X-ray emission from nuclei.}
\label{tab:cores}
\begin{center}
\scriptsize
\begin{tabular}{|lrrrrr|}
\hline
3CR  & Net & F$_{0.5-1~keV}^*$ & F$_{1-2~keV}^*$ & F$_{2-7~keV}^*$ & F$_{0.5-7~keV}^*$ \\
name & counts & (cgs)                 & (cgs)           & (cgs)           & (cgs)   \\
\hline 
\noalign{\smallskip}
91 & 1938.5 (44.0) & 2.5 (0.5) & 21.5 (0.8) & 92.4 (2.7) & 116.4 (2.8) \\
131 & 62.4 (8.0) & -- & 0.2 (0.1) & 3.8 (0.5) & 3.9 (0.5)\\
158 & 288.4 (17.0) & 0.5 (0.2) & 2.8 (0.3) & 9.2 (0.7) & 12.5 (0.8) \\
390 & 1068.5 (32.7) & 4.1(0.6) & 12.7 (0.6) & 40.2 (1.7) & 57.0 (1.9)\\
409 & 596.5 (24.4) & 0.3 (0.2) & 3.6 (0.3) & 35.3 (1.6) & 39.2 (1.7) \\
428 & 557.5 (23.6) & 0.2 (0.1) & 3.9 (0.3) & 28.5 (1.4) & 32.6 (1.4) \\
454.2 & 33.5 (5.8) & -- & 0.1 (0.1) & 2.1 (0.4) & 2.2 (0.4)\\
\noalign{\smallskip}
\hline
\end{tabular}\\
\end{center}
Col. (1): 3CR name.
Col. (2): Background-subtracted number of photons within a circle of radius $r=2$\arcsec\ in the 0.5 - 7 keV band. 
Col. (3): Measured X-ray flux between 0.5 and 1 keV.
Col. (4): Measured X-ray flux between 1 and 2 keV.
Col. (5): Measured X-ray flux between 2 and 7 keV.
Col. (6): Measured X-ray flux between 0.5 and 7 keV.
Note:\\
($^*$) Fluxes are given in units of 10$^{-14}$erg~cm$^{-2}$~s$^{-1}$ and 1$\sigma$ uncertainties in the number of photons computed assuming Poisson statistics are given in parenthesis.
The uncertainties on the flux measurements are computed as described in \S~\ref{sec:obs}. Fluxes were not corrected for Galactic absorption and were computed adopting the same X-ray photometric measurements of \citet{2015ApJS..220....5M}.\\

\end{table*}

\begin{table*} 
	\caption{Results of nuclear X-ray spectral analysis}
	\label{tab:xrayspec}
	\begin{center}
		\begin{tabular}{|lrrrrrr|}
			\hline
			3CR name & z & $N_{H,int}$ & \(\alpha_X\) & $\nu$ & $\chi^2_\nu$ & Luminosity  \\
			&        & (10$^{22}$cm$^{-2}$) &  &  &  & ergs$^{-1}$  \\ 
			\hline 
			\noalign{\smallskip}
			91 & 0.23 & ${0.33}^{+0.25}_{-0.23}$ & 0.8* & 25 & 0.91 & ${1.37}^{+0.12}_{-0.11}\times{10}^{44}$   \\
			91 & 0.23 & 0* & $0.50\pm0.12$ & 25 & 0.77 &  ${1.23}^{+0.22}_{+0.18}\times{10}^{44}$  \\
			91 & 0.19 & 0.31$^{+0.23}_{-0.21}$ & 0.8* & 25 & 0.91 & ${8.87}^{+0.72}_{-0.78}\times{10}^{43}$     \\
			91 & 0.19 & 0* & 0.50$\pm$0.12 & 25 & 0.77 &  ${8.07}^{+1.50}_{-1.22}\times{10}^{43}$  \\
			390 & 0.41 & ${0.50}_{-0.20}^{+0.18}$ & 0.8* & 45 & 0.86 & ${6.03}^{+0.40}_{-0.33}\times{10}^{44}$    \\
			390 & 0.41 & 0* & 0.50$\pm$0.08 & 45 & 0.75 & ${5.45}^{+0.60}_{-0.62}\times{10}^{44}$  \\
			390 & 0.30  & ${0.40}^{+0.16}_{-0.15}$ & 0.8*  & 45 & 0.86 & ${2.89}^{+0.18}_{-0.17}\times{10}^{44}$  \\
			390 & 0.30 & 0* & ${0.50}\pm0.08$ & 45 & 0.75 & ${2.62}^{+0.27}_{-0.25}\times{10}^{44}$ \\
			409 & 1.04 & ${14.4}^{+1.74}_{-1.56}$ & 0.8* & 25 & 0.71 & ${7.51}^{+0.65}_{-0.53}\times{10}^{45}$ \\
			409 & 1.04 & ${11.19}^{+3.00}_{-2.77}$ & ${0.45}^{+0.28}_{-0.27}$ & 24 & 0.68 & ${6.22}^{+3.28}_{-2.71}\times{10}^{45}$ \\
			409 & 0.55 & ${6.80}^{+0.76}_{-0.70}$ & 0.8*& 25 & 0.68 & ${1.40}^{+0.10}_{-0.11}\times{10}^{45}$ \\
			409 & 0.55 & ${5.66}^{+1.51}_{-1.41}$ & ${0.53}^{+0.31}_{-0.30}$ & 24 & 0.68 & ${1.21}^{+0.64}_{-0.52}\times{10}^{45}$ \\
			428 & 2.38 & ${23.37}^{+5.96}_{-5.64}$ & 0.8* & 23 & 0.64 & ${4.32}_{-0.39}^{+0.40}\times{10}^{46}$ \\
			428 & 2.38 &  ${19.89}^{+16.65}_{-14.81}$ & $0.71^{+0.38}_{-0.36}$ & 22 & 0.67 & ${4.09}^{+2.66}_{-2.14}\times{10}^{46}$ \\
			428 & 2.27 & ${21.49}^{+5.52}_{-5.20}$ & 0.8* & 23 & 0.63 & ${3.86}^{+0.34}_{0.34}\times{10}^{46}$ \\
			428 & 2.27 & ${18.37}^{+14.92}_{-13.16}$ & ${0.71}^{+0.36}_{-0.34}$ & 22 & 0.66 & ${3.71}^{+2.49}_{-1.94}\times{10}^{46}$ \\

			\noalign{\smallskip}
			\hline
		\end{tabular}\\
	\end{center}
	Col. (1): The 3CR name.
	Col. (2): Photometric redshifts, obtained from WISE magnitudes in 3.4 and 4.6 $\mu$m filters as described in the text, used for the spectral analysis.
	Col. (3): Intrinsic Absorption, $N_{H,int}$ , as used in the spectral model.
	Col. (4): X-ray Spectral index.
	Col. (5): Degree of freedom.
	Col. (6): Reduced statistic.
	Col. (7): Luminosity of the nucleus, obtained in \textsc{Sherpa} with the function sample\_flux.
	Parameters fixed in the spectral fitting are indicated with *.

\end{table*}

\appendix

\section{A: Radio contours of the sources}
\label{cont}

For all the 3CR sources in our sample, radio contours are shown here. For each source, we have specified the band and configuration at which the observation was performed: L band (1.5 GHz), C band (6 GHz) and X band (10 GHz), in A (maximum baseline, B$_{max}$, equal to 35 km), B (B$_{max}$=10 km), C (B$_{max}$=3.5 km) or D (B$_{max}$=1 km) configuration.
In the bottom left corner, the clean beam is shown as a black filled ellipse. All the details about the single observations are reported in Table \ref{tab:radio_obs}.

\begin{figure*}
\centering
\includegraphics[scale=0.3]{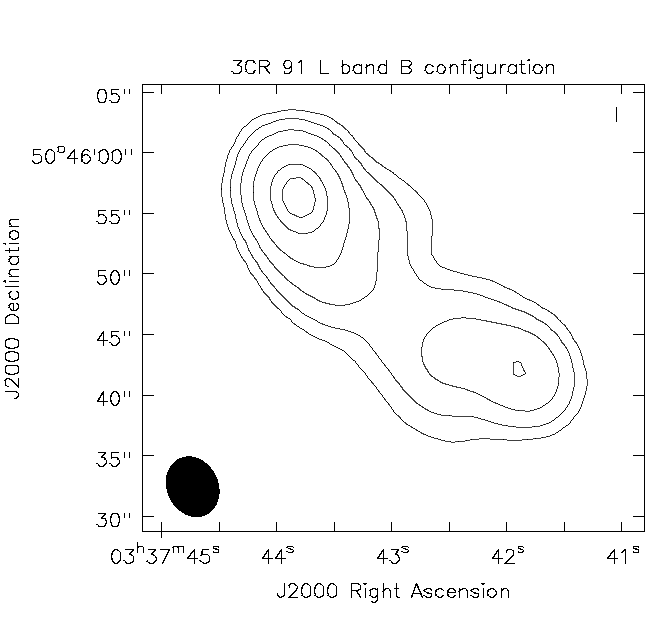}
\includegraphics[scale=0.3]{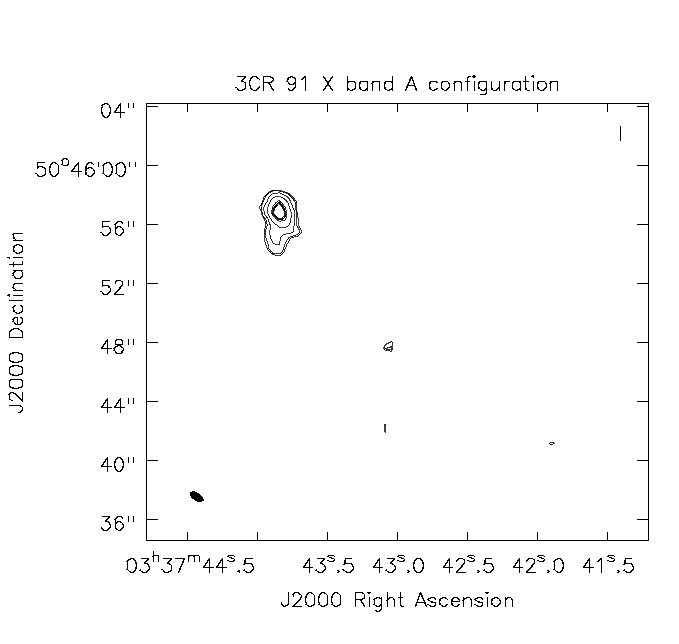}
\caption {Radio VLA contours of 3CR\,91. The name of the source, the observing band, and array configuration are reported on the top of each panel. The clean beam is shown as a black filled ellipse in the bottom left of the image. Radio contours are reported in the last column in Table \ref{tab:radio_obs}. On the top of each panel, name, band and configuration are reported.}
\label{3c91_radio}
\end{figure*}

\begin{figure*}
\centering
\includegraphics[scale=0.35]{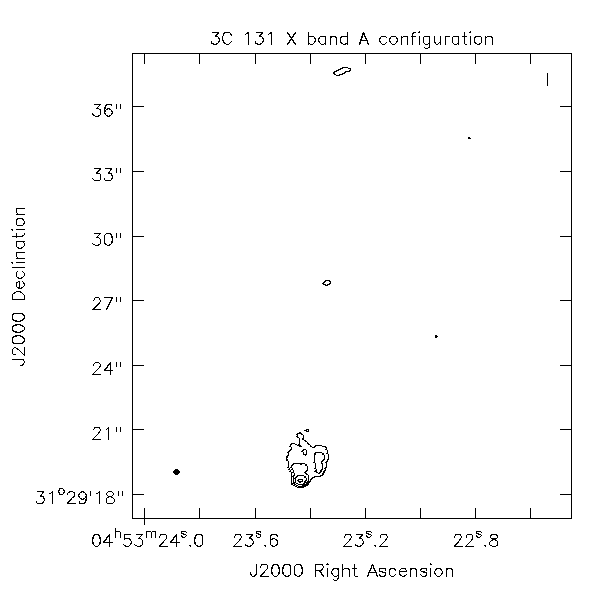}
\includegraphics[scale=0.3]{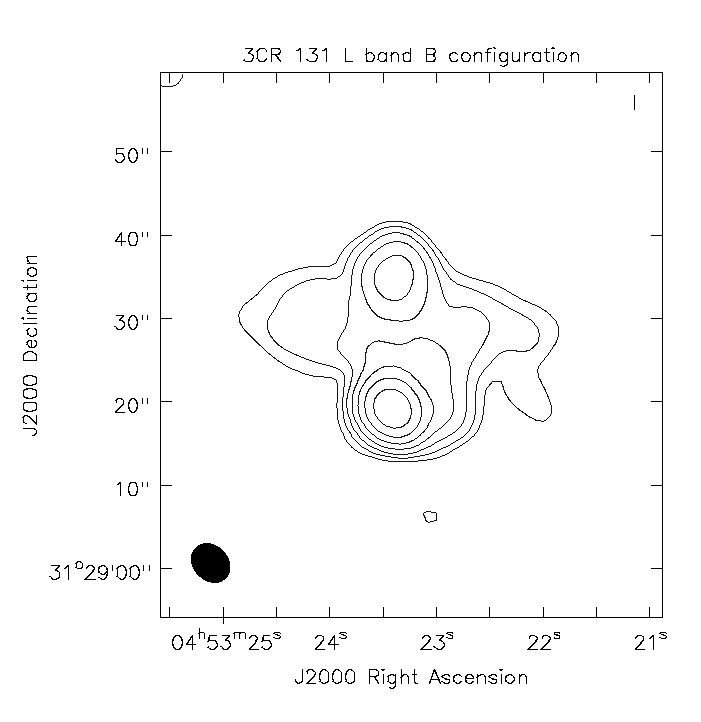}
\caption {Radio VLA contours of 3CR\,131. The name of the source, the observing band, and array configuration are reported on the top of each panel. The clean beam is shown as a black filled ellipse in the bottom left of the image. Radio contours are reported in the last column in Table \ref{tab:radio_obs}.}
\label{3c131_radio}
\end{figure*}

\begin{figure*}
\centering
\includegraphics[scale=0.3]{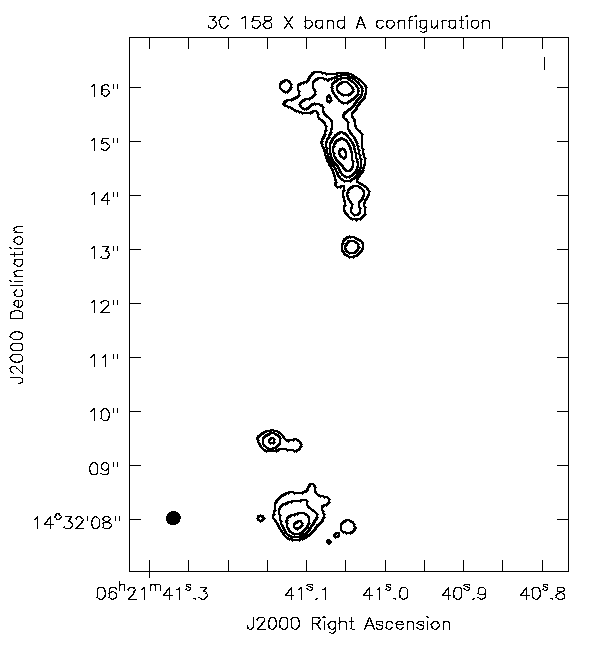}
\includegraphics[scale=0.28]{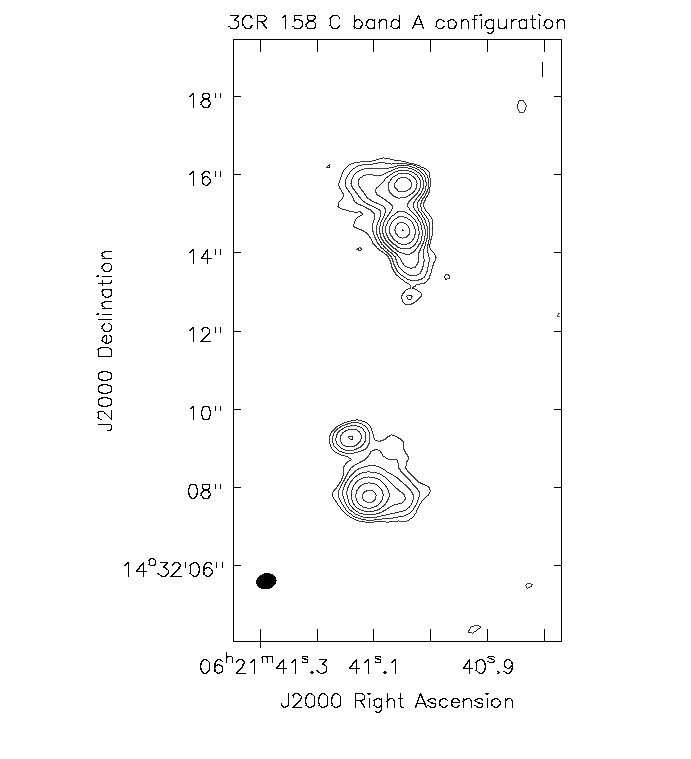}
\includegraphics[scale=0.25]{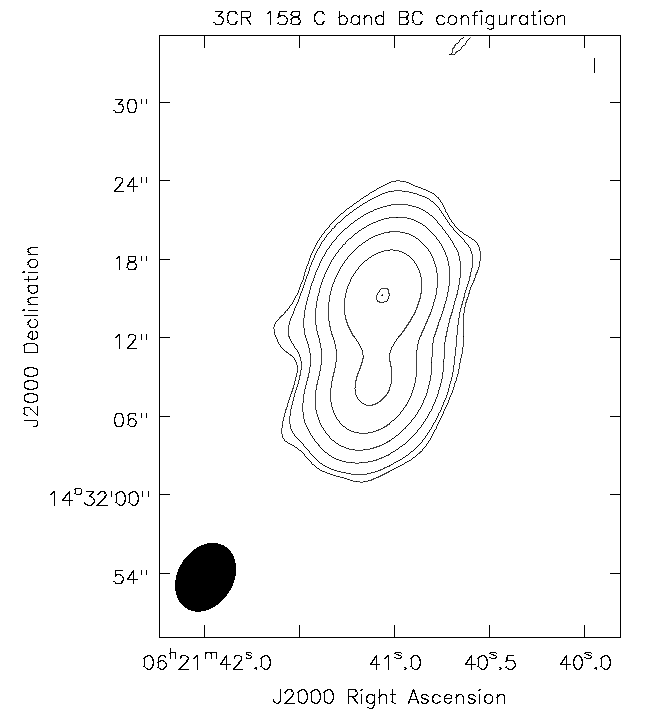}
\caption {Radio VLA contours of 3CR\,158. The name of the source, the observing band, and array configuration are reported on the top of each panel. The clean beam is shown as a black filled ellipse in the bottom left of the image. Radio contours are reported in the last column in Table \ref{tab:radio_obs}.}
\label{3c158_radio}
\end{figure*}

\begin{figure*}
\centering
\includegraphics[scale=0.3]{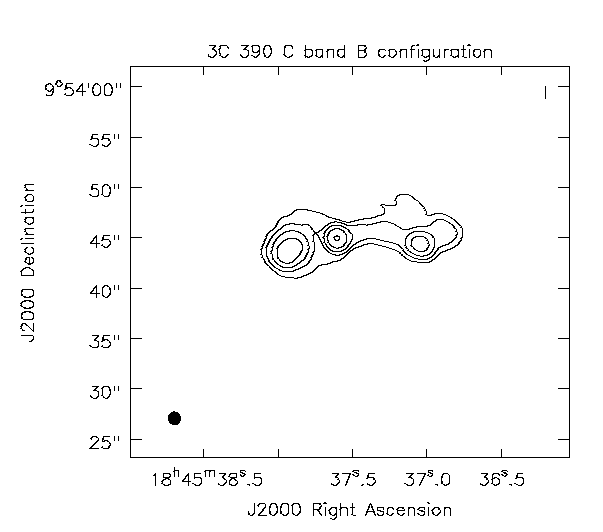}
\includegraphics[scale=0.28]{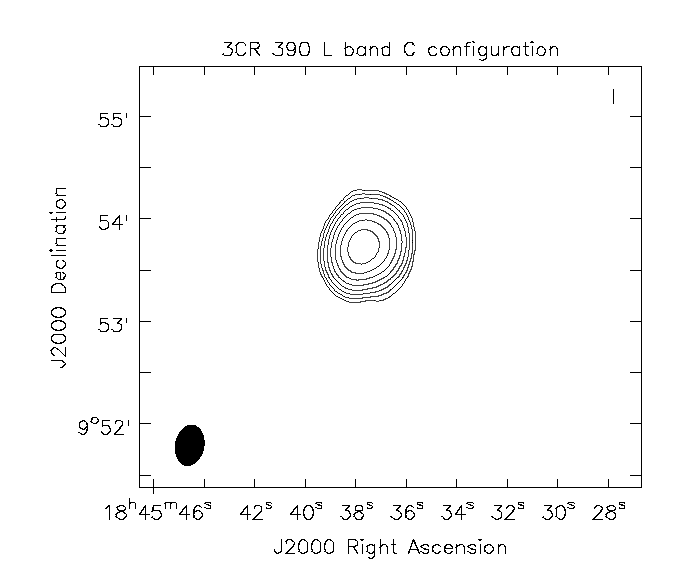}
\includegraphics[scale=0.25]{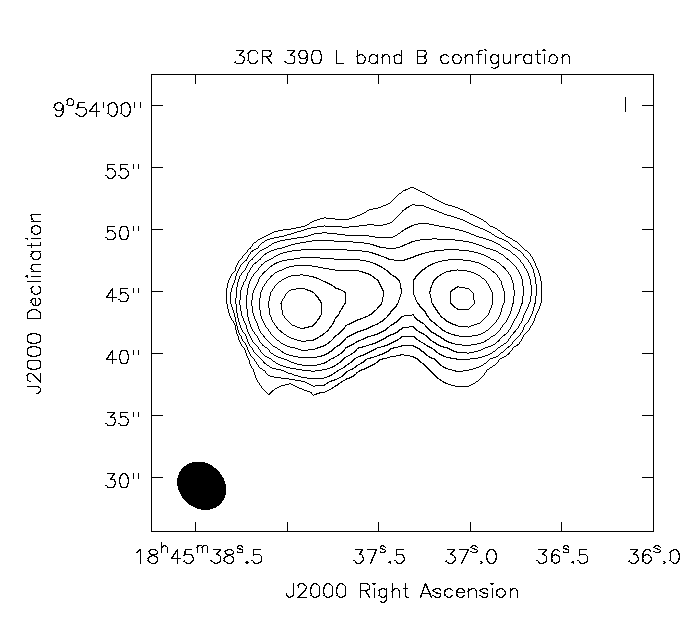}
\caption {Radio VLA contours of 3CR\,390. The name of the source, the observing band, and array configuration are reported on the top of each panel. The clean beam is shown as a black filled ellipse in the bottom left of the image. Radio contours are reported in the last column in Table \ref{tab:radio_obs}.}
\label{3c390_radio}
\end{figure*}

\begin{figure*}
\centering
\includegraphics[scale=0.3]{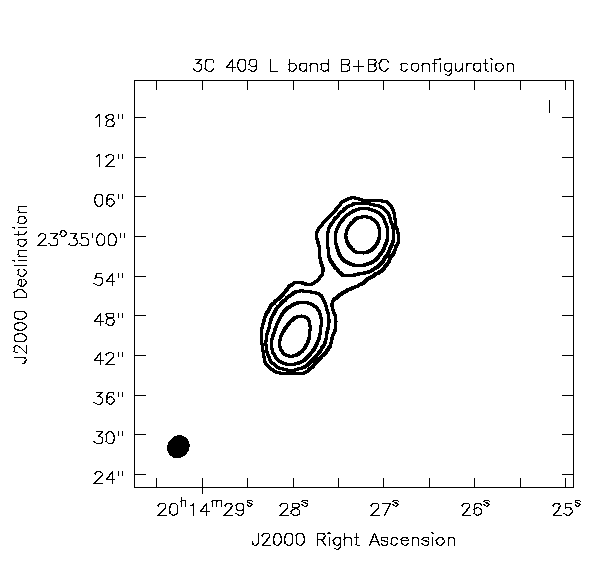}
\caption {Radio VLA contours of 3CR\,409. This map has been obtained merging two observations in B and BC configuration, respectively, as described in Sect. \ref{sec:results}. The name of the source, the observing band, and array configuration are reported on the top of each panel. The clean beam is shown as a black filled ellipse in the bottom left of the image. Radio contours are reported in the last column in Table \ref{tab:radio_obs}.}
\label{3c409}
\end{figure*}

\begin{figure*}
\centering
\includegraphics[scale=0.35]{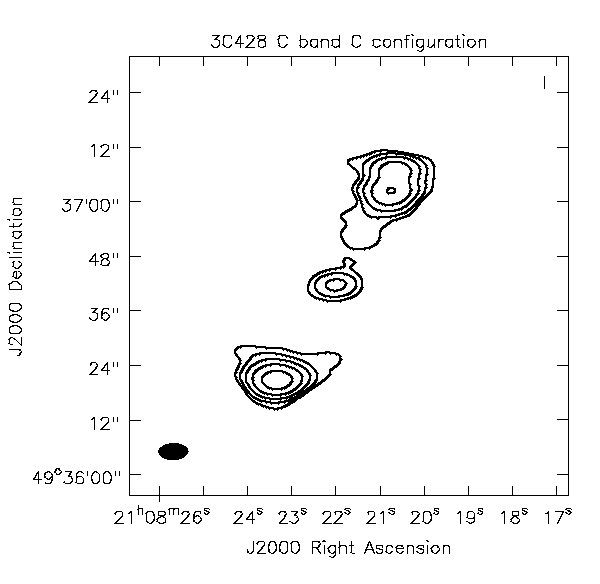}
\includegraphics[scale=0.3]{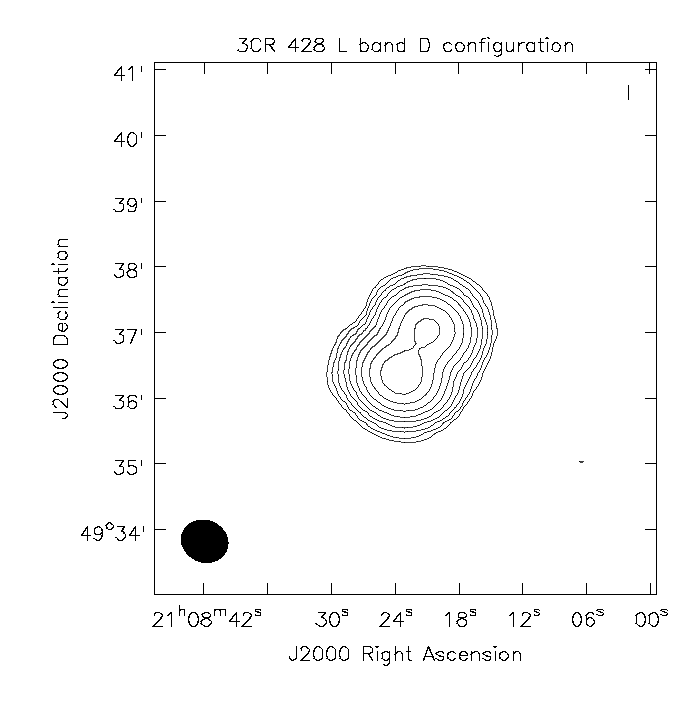}
\caption {Radio VLA contours of 3CR\,428. The name of the source, the observing band, and array configuration are reported on the top of each panel. The clean beam is shown as a black filled ellipse in the bottom left of the image. Radio contours are reported in the last column in Table \ref{tab:radio_obs}.}
\label{3c428}
\end{figure*}

\begin{figure*}
\centering
\includegraphics[scale=0.4]{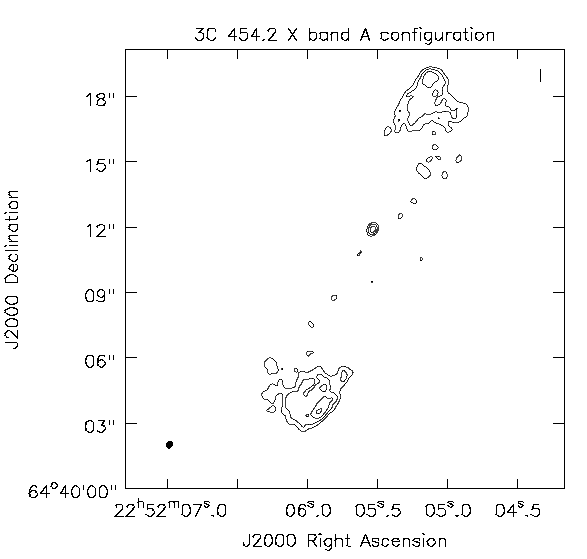}
\includegraphics[scale=0.35]{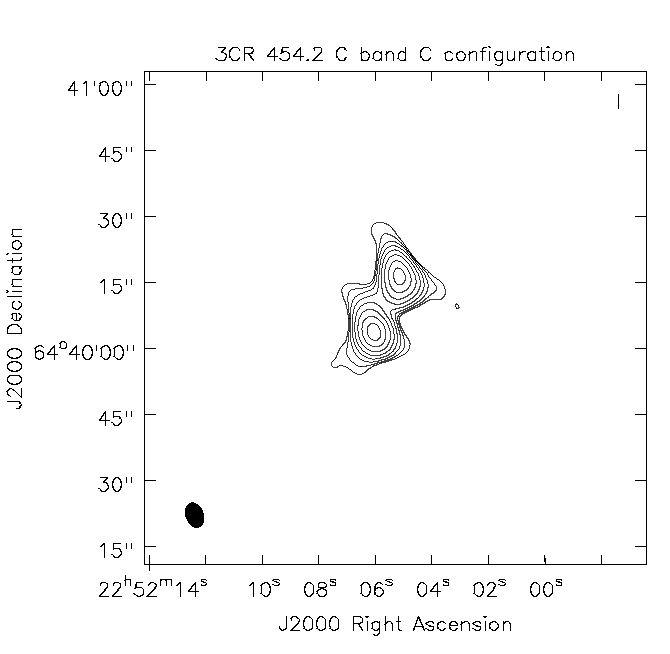}
\caption {Radio VLA contours of 3CR\,454.2. The name of the source, the observing band, and array configuration are reported on the top of each panel. The clean beam is shown as a black filled ellipse in the bottom left of the image. Radio contours are reported in the last column in Table \ref{tab:radio_obs}.}
\label{fig:radio_fig}
\end{figure*}


\end{document}